\documentclass[letter]{aa}  

\usepackage{graphicx}
\usepackage{txfonts}
\usepackage{lipsum}
\usepackage{subcaption}         
\usepackage{lscape}             
\usepackage{placeins}           
                                
\usepackage{hyperref}
\usepackage{ulem}
\usepackage{orcidlink}
\usepackage{multirow}

\begin{document}

   \title{The variability angular diameter distance and the intrinsic brightness temperature of active galactic nuclei}


   \titlerunning{The intrinsic brightness temperature of active galactic nuclei}

   \author{Whee Yeon Cheong\inst{1,2}\fnmsep\thanks{\email{wheeyeon@kasi.re.kr}}$^{\orcidlink{0009-0002-1871-5824}}$
        \and Sang-Sung Lee\inst{1,2}\fnmsep\thanks{\email{sslee@kasi.re.kr}}$^{\orcidlink{0000-0002-6269-594X}}$
        \and Chanwoo Song\inst{1,2}
        \and Jeffrey Hodgson\inst{3}$^{\orcidlink{0000-0001-6094-9291}}$
        \and Sanghyun Kim\inst{2}$^{\orcidlink{0000-0001-7556-8504}}$
        \and Hyeon-Woo Jeong\inst{1,2}$^{\orcidlink{0009-0005-7629-8450}}$
        \and Young-Bin Shin\inst{1,2}
        \and Sincheol Kang\inst{2}$^{\orcidlink{0000-0002-0112-4836}}$
        }

   \institute{University of Science and Technology, 217 Gajeong-ro, Yuseong-gu, Daejeon 34113, Republic of Korea
            \and Korea Astronomy and Space Science Institute, 776 Daedeok-daero, Yuseong-gu, Daejeon 34055, Republic of Korea
            \and Department of Physics and Astronomy, Sejong University, 209 Neungdong-ro, Gwangjin-gu, Seoul, Republic of Korea}

   \date{Received September 30, 20XX}

 
  \abstract
   {It has recently been suggested that angular diameter distances derived from comparing the variability timescales of blazars to angular size measurements with very long baseline interferometry (VLBI) may provide an alternative method to study the cosmological evolution of the Universe. Once the intrinsic brightness temperature ($T_{\rm int}$) is known, the angular diameter distance may be found without knowledge of the relativistic Doppler factor, opening up the possibility of a single rung distance measurement method from low $(z_{\rm cos}\ll1)$ to high $(z_{\rm cos}>4)$ redshifts. Previous studies have found $T_{\rm int}\approx10^{10}$---$10^{11}$~K, with a potential frequency dependence.}
   {We aim to verify whether the variability-based estimates of the intrinsic brightness temperature of multiple active galactic nuclei (AGNs) converges to a common value. We also investigate whether the intrinsic brightness temperature changes as a function of frequency.}
   {We estimated the $T_{\rm int}$ of AGNs based on the flux variability of the radio cores of their jets. We utilized radio core light curves and size measurements of 75 sources at 15 GHz and of 
   37 sources at 43 GHz. We also derived $T_{\rm int}$ from a population study of the brightness temperatures of VLBI cores using VLBI survey data of more than $100$ sources at 24, 43, and 86~GHz.}
   {Radio core variability-based estimates of $T_{\rm int}$ constrain upper limits of $\log_{10}T_{\rm int}\textrm{~[K]}<11.56$ at 15~GHz and $\log_{10}T_{\rm int}\textrm{~[K]}<11.65$ at 43~GHz under a certain set of geometric assumptions. The population analysis suggests lower limits of $\log_{10}T_{\rm int}\textrm{~[K]}>9.7$, $9.1$, and $9.3$ respectively at 24, 43, and 86~GHz. Even with monthly observations, variability-based estimates of $T_{\rm int}$ appear to be cadence-limited.}
   {Methods used to constrain $T_{\rm int}$ are more uncertain than previously thought. However, with improved datasets, the estimates should converge.}


   \maketitle

\section{Introduction}
\makeatletter
\@ifpackageloaded{linenoaa}{\nolinenumbers}{}
\makeatother
With the continued presence of the Hubble tension \citep{2020A&A...641A...6P,2022ApJ...934L...7R} and hints of evolving dark energy \citep{2025JCAP...02..021A}, it is of ever-growing importance to cross evaluate the known cosmological parameters with multiple independent methods. One such method uses the ratio of the observed angular size and the causality-limited size to determine the angular diameter distance to a variable source \citep{2001A&A...366.1061W,2020MNRAS.495L..27H} . The idea stems from the assumption that the variability timescale, $\Delta t$, of a source connects to the physical size of the emission region as
\begin{equation}
    a=fc\Delta t,
\end{equation}
where $a$ is the radial size of the source (if $f=0.5$, $c\Delta t$ corresponds to the diameter), $c$ is the speed of light, and $f$ is some correction factor equating the causality size ($c\Delta t$) to the physical linear size. A correlation between $a$ (or the angular radial size $\theta_{\rm a}$) and $\Delta t$ has been found in active galactic nuclei (AGNs) \citep[see e.g.,][]{2023MNRAS.525.5105H}, supporting our assumptions. From the definition of the angular diameter distance, we then have \citep{2020MNRAS.495L..27H}
\begin{equation}
    D_{\rm A} = \frac{a}{\theta_{\rm a}} = \frac{fc\delta\Delta t^{\rm o}_{\nu^{\rm o}}}{\left[1+z_{\rm cos}\right]\theta_{\rm a}},
    \label{eq:da_eq_1}
\end{equation}
where we have taken (here, as well as in the remainder of the text) $f=1$, $\Delta t^{\rm o}_{\nu^{\rm o}}$ as the variability timescale ($\Delta t^{\rm o}$) measured in the observer frame, $\mathcal{F}^{\rm o}$ (denoted with the superscript ``o''), at frequency $\nu^{\rm o}$; $z_{\rm cos}$ is the cosmological redshift of the source; and $\theta_{\rm a}$ is the angular size.\\
\indent{}The calculation of $D_{\rm A}$ using Equation~(\ref{eq:da_eq_1}) requires knowledge of the Doppler factor, $\delta$, in order to correct for potential unknown special relativistic effects. \citet{2023MNRAS.521L..44H} assumed a common maximum intrinsic brightness temperature ($T_{\rm int}$) to remove an explicit dependence on $\delta$. It was found that
\begin{equation}
    D_{\rm A}=\frac{2\ln{2}c^3\Delta t^{\rm o}_{\nu^{\rm o}}S^{\rm o}_{\nu^{\rm o}}}{\pi k_{\rm B}T_{\rm int,\nu^{\rm e}}\nu^{\textrm{o}2}\theta_{\rm vlb}^3},
    \label{eq:hod2023_orig_v1}
\end{equation}
where $\nu^{\rm e}=\nu^{\rm o}\left[1+z_{\rm cos}\right]/\delta$ is the frequency of the observed photon in the emission frame (denoted with the superscript ``e''). The very long baseline interferometry (VLBI) measured angular size $\theta_{\rm vlb}=B\theta_{\rm FWHM}$ is the full width at half maximum (FWHM) of the fit Gaussian component ($\theta_{\rm FWHM}$) multiplied by some correction factor, $B$, that accounts for the source geometry. Typically, Gaussians are fit to visibilities to determine component angular sizes. However, a Gaussian is not a realistic approximation of the true source geometry. In principle, with high quality observations, the true source geometry may be measured \citep{2015A&A...574A..84L}. \\
\indent{}The assumption of a common value of $T_{\rm int}$ over different sources and frequencies may not be correct. Previous studies \citep[e.g.,][]{2018ApJ...866..137L,2006ApJ...642L.115H,2021ApJ...923...67H,2013JKAS...46..243L,2014JKAS...47..303L,2000A&A...364..391L,2019A&A...622A..92N} using various methods have found values of $T_{\rm int}$ of $\sim$$10^{10}$---$10^{11}$~K. \citet{2014JKAS...47..303L} assumed a population model of AGNs to suggest a possible frequency dependence of the value of $T_{\rm int}$, peaking at $\nu^{\rm o}\sim10$~GHz. Such a frequency dependence, if left unaccounted for, will introduce a redshift-dependent bias into the cosmological measurements using the $T_{\rm int}$ assumption. By switching $T_{\rm int,\nu^{\rm e}}$ and $D_{\rm A}$ in Equation~(\ref{eq:hod2023_orig_v1}), we can estimate the $T_{\rm int}$ for an individual source with constrained values of $\Delta t$, $S_{\nu}$, $\theta_{\rm vlb}$, and $z_{\rm cos}$ and an assumed cosmology. The purpose of this paper is to a) utilize a refined version of Equation~(\ref{eq:hod2023_orig_v1}) to verify whether the $T_{\rm int}$ calculated for multiple sources does indeed converge at a certain value and b) with additional $T_{\rm vlb}$ versus apparent speed ($\beta_{\rm app}$) analysis at 24, 43, and 86~GHz, investigate whether there is indeed a frequency dependence of $T_{\rm int}$.
\section{Estimating the intrinsic brightness temperature}
The intrinsic brightness temperature, $T_{\rm int}$, of the emission region is difficult to obtain, as the observed brightness temperature is affected by $\delta$. While $\delta$ may be estimated under various assumptions \citep[e.g.,][]{2017ApJ...846...98J,2009A&A...494..527H,2017MNRAS.466.4625L,1993ApJ...407...65G,1994ApJ...426...51R, 1996ApJ...461..600G}, it is of interest to derive $T_{\rm int}$ independently of $\delta$. We utilized the following two methods to do so.

\subsection{Variability-based estimate}\label{sec:Tint_var}
The (causality) constrained physical extent of the emission region and the definition of the angular diameter distance $D_{\rm A}$ may be used to obtain Equation~(\ref{eq:da_eq_1}). Parameters defined in the host-galaxy frame, $\mathcal{F}^{\rm a}$, are denoted with the superscript ``a" (i.e., $\nu^{\rm o}=\nu^{\rm a}/\left[1+z_{\rm cos}\right]=\delta\nu^{\rm e}/\left[1+z_{\rm cos}\right]$).
Following \citet{2023MNRAS.521L..44H}, we substituted $\delta=T^{\rm a}_{\rm vlb, \nu^{\rm a}}/T_{\rm int,\nu^{\rm e}}$ to find
\begin{equation}
    D_{\rm A}
    =\frac{c\Delta t^{\rm o}_{\nu^{\rm o}}}{\left[1+z_{\rm cos}\right]\theta_{\rm a}}\frac{T^{\rm a}_{\rm vlb,\nu^{\rm a}}}{T_{\rm int,\nu^{\rm e}}}
    =\frac{c\Delta t^{\rm o}_{\nu^{\rm o}}}{\theta_{\rm a}}\frac{T^{\rm o}_{\rm vlb,\nu^{\rm o}}}{T_{\rm int,\nu^{\rm e}}},
    \label{eq:main_da_eq_wyc_v2_0}
\end{equation}
where $T_{\rm int,\nu}$ is the intrinsic brightness temperature of the source and $T_{\rm vlb,\nu}$ is the VLBI-measured brightness temperature.\\
\indent{}Proceeding further, we assumed either a uniform circular disk or an optically thin spherical geometry as more realistic models for the emission region. Taking $\theta_{\rm FWHM}$ as the angular FWHM of the model-fit Gaussian component, the total radial angular size of a uniform disk is $\theta_{\rm a}\approx0.8\theta_{\rm FWHM}$ and that of a sphere is $\theta_{\rm a}\approx0.9\theta_{\rm FWHM}$ \citep[e.g.,][]{1995ASPC...82..267P}. Based on these approximations, we rewrote Equation~(\ref{eq:hod2023_orig_v1}) to solve for $T_{\rm int,\nu^{\rm e}}$ as 
\begin{equation}
    T_{\rm int,\nu^{\rm e}}
    =X\frac{2\ln{2}c^3\Delta t^{\rm o}_{\nu^{\rm o}}S^{\rm o}_{\nu^{\rm o}}}{\pi k_{\rm B}D_{\rm A}\nu^{\textrm{o}2}\theta_{\rm FWHM}^3},
    \label{eq:wyc_tbint_re_0}
\end{equation}
where X equals 1 for a Gaussian, $\approx0.70$ for a disk, and $\approx0.74$ for a sphere (see Appendix~\ref{sec:app_eqderv} for details). For a source with a known characteristic variability timescale, flux density, size, and (angular diameter) distance, we can evaluate the intrinsic brightness temperature, $T_{\rm int,\nu^{\rm e}}$, in the emission frame, $\mathcal{F}^{\rm e}$, without prior knowledge of the Doppler factor, $\delta$. It should be noted that this is equivalent to estimating $T_{\rm int,\nu}$ from $T_{\rm vlb,\nu}$ and the variability brightness temperature $\left(T_{\rm var,\nu}\right)$ as $T_{\rm int,\nu^{\rm e}}=\sqrt{T^{\rm o 3}_{\rm vlb,\nu^{\rm o}}/T^{\rm o}_{\rm var,\nu^{\rm o}}}$. It is assumed that $T_{\rm vlb,\nu}$ and $T_{\rm var,\nu}$ are calculated for the same emission region geometry and that the same flux density value, $S^{\rm o}_{\nu^{\rm o}}$, is used. In \citet{2023MNRAS.521L..44H}, assuming a Gaussian geometry, $T_{\rm int,\nu^{\rm e}}$ was found to be of $\sim$$4\times10^{11}$~K. We find that $T_{\rm int,\nu^{\rm e}}$ estimated with Equation~(\ref{eq:wyc_tbint_re_0}) is $\sim4\times10^{10}$~K (see Appendix~\ref{sec:app_3c84_c3}). We emphasize that inconsistent use of geometric assumptions can lead to differences of approximately an order of magnitude in the derived quantities.
\subsection{Population study with $T_{\rm vlb,\nu}$ and $\beta_{\rm app}$}\label{sec:Tint_pop}
A different method of estimating $T_{\rm int}$ utilizes the distribution of the apparent speeds, $\beta_{\rm app}$, of the jet with respect to $T^{\rm a}_{\rm vlb,\nu^{\rm a}}$ \citep{2006ApJ...642L.115H,2021ApJ...923...67H}. From special relativity, we have $\delta=\left[\gamma_{\rm j}\left(1-\beta_{\rm j}\cos{\theta_{\rm j}}\right)\right]^{-1}$ and $\beta_{\rm app}=\beta_{\rm j}\sin{\theta_{\rm j}}/\left(1-\beta_{\rm j}\cos{\theta_{\rm j}}\right)$. Here, $\gamma_{\rm j}=\left(1-\beta_{\rm j}^{2}\right)^{-1/2}$ is the bulk Lorentz factor of the jet, and $\theta_{\rm j}$ is the viewing angle (i.e., the angle between the jet axis and the line of sight). For a given $\gamma_{\rm j}$, we can estimate $\beta_{\rm app}$ as 
\begin{equation}
    \beta_{\rm app}
    =\left[\frac{2\gamma_{\rm j}T_{\rm vlb,\nu^{\rm a}}^{\rm a}}{T_{\rm int,\nu^{\rm e}}}-\left(\frac{T_{\rm vlb,\nu^{\rm a}}^{\rm a}}{T_{\rm int,\nu^{\rm e}}}\right)^2-1\right]^{\frac{1}{2}}
    .
    \label{eq:obs_bapp_Tvlb}
\end{equation}
For a source with given $T_{\rm int,\nu^{\rm e}}$ and $\gamma_{\rm j}$, the observed $T_{\rm vlb,\nu^{\rm a}}^{\rm a}$ and $\beta_{\rm app}$ varies due to $\theta_{\rm j}$. The maximum observable apparent speed ($\beta_{\rm max}$) of
\begin{equation}
    \beta_{\rm max}=\left[\left(\frac{T_{\rm vlb,\nu^{\rm a}}^{\rm a}}{T_{\rm int,\nu^{\rm e}}}\right)^2-1\right]^{\frac{1}{2}},
    T_{\rm int,\nu^{\rm e}}\leq T_{\rm vlb,\nu^{\rm a}}^{\rm a}
    \label{eq:max_bapp_Tvlb}
\end{equation}
is obtained when the jet is viewed at the critical angle $\theta_{\rm j}=\theta_{\rm c}\equiv\arccos{\beta_{\rm j}}$. For a flux-limited sample of sources with a common $T_{\rm int,\nu^{\rm e}}$, approximately $75\%$ of the sources are expected to have viewing angles of $\theta_{\rm j}<\theta_{\rm c}$ \citep{2006ApJ...642L.115H,2013JKAS...46..243L}. This manifests as $75\%$ of sources lying below the $\beta_{\rm max}$ line in the $\beta_{\rm app}$ versus $T_{\rm vlb,\nu^{\rm a}}^{\rm a}$ parameter space. Under these assumptions, all $\beta_{\rm app}$ measurements should lie under the envelope provided by the maximum observed $\beta_{\rm app}$ (solid line in Figure~\ref{fig:Tint_pop_per_source}). Therefore, we may constrain the value of $T_{\rm int,\nu^{\rm e}}$ using a flux-limited sample of AGNs with known $\beta_{\rm app}$ and $T_{\rm vlb,\nu^{\rm a}}^{\rm a}$ \citep{2006ApJ...642L.115H,2021ApJ...923...67H,2013JKAS...46..243L,2014JKAS...47..303L}.
\subsection{Data analysis}
For the variability-based analysis (Section~\ref{sec:Tint_var}), we applied Equation~(\ref{eq:wyc_tbint_re_0}) to the archival Very Long Baseline Array (VLBA) data from the MOJAVE program\footnote{\url{https://www.cv.nrao.edu/MOJAVE/}} at 15~GHz \citep{2018ApJS..234...12L} and the Boston University VLBA-BU-BLAZAR 43~GHz monitoring program\footnote{\url{https://www.bu.edu/blazars/BEAM-ME.html}} \citep{2017ApJ...846...98J, 2022ApJS..260...12W}. We collected multi-epoch core fluxes and sizes for multiple sources as a result of model-fitting multiple Gaussian components to the observed visibilities. At 15~GHz, we utilized the tabulated data published in \citet{2021ApJ...923...67H}. At 43~GHz, we combined the data from 2007 June to 2012 December in \citet{2017ApJ...846...98J} and from 2013 January to 2018 December in \cite{2022ApJS..260...12W}. For each source, we evaluated the variability timescale for each pair of consecutive measurements as 
\begin{equation}
    \tau_{i+1,i}=\frac{t_{i+1}-t_{i}}{\ln{S_{i+1}}-\ln{S_{i}}}. \nonumber
\end{equation}
We used $\tau_{i+1,i}$, along with the flux density and size measurements of the epoch with the larger flux density to calculate $T_{\rm int}$ for each measurement pair. When doing so, we only kept values for which both $\theta_{\rm FWHM,i+1}$ and $\theta_{\rm FWHM,i}$ were marked as resolved in the original publications, and only if $\tau_{i+1,i}>0$ (i.e., the flux density is rising between consecutive measurements). From the constraint on $\tau_{i+1,i}$, we ultimately utilized $S_{i+1}$ and $\theta_{i+1}$ for our flux density and size measurements. 
The two-point estimate of $\tau_{i+1,i}$ relies on significant flux variability measurements with a cadence comparable to, or better than, the variability timescale of the flare in question. Estimates of $\tau_{i+1,i}$ may be biased toward larger values should these conditions not be met. We tested for cadence-limited biases by alternating the constraint on $t_{i+1}-t_{i}$ to be less than 30, 50, 100, and 500~days and unconstrained (see Appendix~\ref{sec:app_estimation_of_Tint} for details).
Uncertainties in the value of $T_{\rm int}$ were determined by drawing $10^{4}$ random values for each measurement of $S_{i}$ and $\theta_{\rm FWHM,i}$ from a normal distribution with the mean and standard deviation set respectively to each measured value and measurement uncertainty. The median, 15.865\%, and 84.135\% percentiles were used to evaluate the value and uncertainties of $\log_{10}T_{\rm int}$ for each source.\\
\indent{}For the population study (Section~\ref{sec:Tint_pop}), we collected the radio core size and flux measurements at multiple frequencies with VLBI observations. At 24~GHz, we used the data from \citet{2023AJ....165..139D}, which contains the model-fitting results of 731 unique sources from 29 observations (from 2015 July to 2018 July) with the VLBA. Of the 731 unique sources, we found 598 sources with constrained $T_{\rm vlb,\nu^{\rm a}}^{\rm a}$ measurements (i.e., with known cosmological redshift and the core component resolved). Of these sources, 199 have $\beta_{\rm app}$ measurements in \citet{2021ApJ...923...30L}. At 43~GHz, we collected the data from the VLBA 43~GHz imaging survey in \citet{2018ApJS..234...17C,2020ApJS..247...57C}. Of the 134 sources, we found 99 sources with $\beta_{\rm app}>0$ from \citet{2021ApJ...923...30L}. At 86~GHz, we combined the data in \citet{2008AJ....136..159L} and \citet{2019A&A...622A..92N}, both obtained with the Global Millimeter VLBI Array. We found a total of 141 unique sources with $\beta_{\rm app}$ measurements from \citet{2021ApJ...923...30L, 2019ApJ...874...43L}. At all frequencies, we used the maximum apparent speed for each source \citep{2021ApJ...923...67H}.
\section{Results}
Values of $T_{\rm int}$ were found in the range of $\log_{10}T_{\rm int}\textrm{~[K]}=10.73^{+0.83}_{-0.99}$ to $11.19^{+0.75}_{-0.80}$ at 15~GHz and $\log_{10}T_{\rm int}\textrm{~[K]}=10.74^{+0.91}_{-1.12}$ to $10.92^{+0.94}_{-0.96}$ at 43~GHz. From the central limit theorem, we expected $T_{\rm int}$ to have a log-normal distribution. We conducted a D’Agostino and Pearson’s test \citep{10.1093/biomet/60.3.613} and a Shapiro-Wilk test \citep{eb32428d-e089-3d0c-8541-5f3e8f273532} to quantify any deviations from normality. The distribution of $\log_{10}T_{\rm int}$ over multiple sources approximately follows a normal distribution at 43~GHz, while we measured a significant deviation from normality at 15~GHz (Appendix~\ref{sec:app_estimation_of_Tint}). 
We also found that the values of $\log_{10}T_{\rm int}$ significantly decrease with tighter constraints on $\left(t_{i+1}-t_{i}\right)$, i.e., $\max\left(t_{i+1}-t_{i}\right)$ from $\infty$ to 30 days. With a sufficient observational cadence, we expected the value of $\log_{10}T_{\rm int}$ to converge on some specific value instead of continuing to decrease with smaller $\max\left(t_{i+1}-t_{i}\right)$. This suggests that the variability-based estimate of $\log_{10}T_{\rm int}$ at both frequencies may be biased toward higher values. Therefore, we placed upper limits of $\log_{10}T_{\rm int}\textrm{~[K]}<11.56$ at 15~GHz and $\log_{10}T_{\rm int}\textrm{~[K]}<11.65$ at 43~GHz, obtained from the 84.135\% percentile value with $t_{i+1}-t_{i}<30$~days.\\
\indent{}From the population analysis of the 24, 43, and 86~GHz data, we found (for a disk-like geometry) $\log_{10}{T_{\rm int,24}\textrm{~[K]}}=9.85^{+0.32}_{-0.12}$, $\log_{10}T_{\rm int,43}\textrm{~[K]}=9.23^{+0.35}_{-0.12}$, and $\log_{10}T_{\rm int,86}~\textrm{[K]}=9.36^{+0.24}_{-0.06}$, respectively. Here, the uncertainties on $T_{\rm int}$ were estimated following \citet{2006ApJ...642L.115H,2013JKAS...46..243L}, where the value of $T_{\rm int}$ with 75\% of sources below the $\beta_{\rm app}$ line ($T_{\rm int}$) was taken as the nominal value, with the values containing 60\% and 80\% of the sources respectively as the upper and lower bounds on $T_{\rm int}$. 
\section{Discussion}
\begin{figure}[t!]
    \centering
    \includegraphics[width=\linewidth]{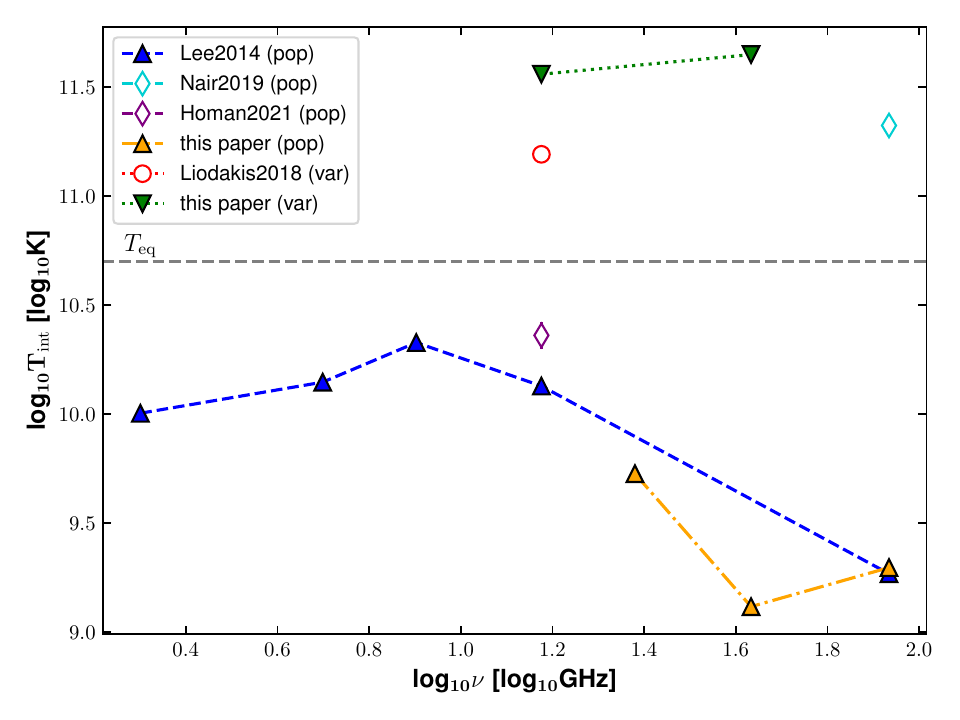}
    \caption{Intrinsic brightness temperature, $T_{\rm int}$, of AGNs obtained at different $\nu^{\rm o}$. Values with the label ``pop'' were obtained from a population analysis of $T_{\rm vlb}$ and $\beta_{\rm app}$. Values with the label ``var'' were obtained utilizing flux density variability. All values were scaled to correspond to a disk source geometry. The dashed horizontal line corresponds to the $T_{\rm eq}\approx5\times10^{10}$~K of \cite{1994ApJ...426...51R}. The values of Lee2014 are from \cite{2014JKAS...47..303L}, Nair2019 from \cite{2019A&A...622A..92N}, Homan2021 from \cite{2021ApJ...923...67H}, and Liodakis2018 from \cite{2018ApJ...866..137L}, and they have been adjusted for the source geometry.}
    \label{fig:multifreqTbint}
\end{figure}

There is a large offset between $T_{\rm int}$ estimated from variability analysis and population analysis, with variability-based estimates resulting in systematically larger values. We have shown that the current data at both 15 and 43~GHz is cadence-limited (see Appendix~\ref{sec:app_estimation_of_Tint} for details). Additionally, the current estimate of $T_{\rm int}$ using Equation~(\ref{eq:wyc_tbint_re_0}) relies on the assumption that $a=c\Delta t$ (i.e., $f=1$). This ``causality-constrained'' linear size should be taken as an upper limit on $a$, with the true value of $a$ being lower. In such a case, Equation~(\ref{eq:wyc_tbint_re_0}) may overestimate $T_{\rm int}$. Both causality arguments and data limitations suggest that the variability-based estimates of $T_{\rm int}$ should be considered as upper limits.\\
\indent{}The population analysis method of \citet{2006ApJ...642L.115H} relies on brightness temperature measurements of a complete, flux-limited sample of sources, with accompanying measurements of the apparent jet speeds. \citet{2021ApJ...923...67H} utilized the MOJAVE 1.5~Jy quarter-century sample with the population model of \citet{2019ApJ...874...43L} to find (once rescaled to a disk) $\log_{10}T_{\rm int,15}\textrm{~[K]}=10.36_{-0.06}^{+0.06}$. This is comparable to \citet{2014JKAS...47..303L}. The samples at the other frequencies are not flux complete, which may bias the results. For some sources, only lower limits on $T_{\rm vlb}$ (due to unresolved cores) are available. Including such lower limits in the analysis does not significantly alter the final value. As described in Section~\ref{sec:Tint_pop},  all $\beta_{\rm app}$ measurements should lie under the envelope (i.e., Equation~(\ref{eq:obs_bapp_Tvlb}) from maximum observed $\beta_{\rm app}$). However, we find that a large number of sources lie outside of the envelope, with most of the outliers having a larger $T_{\rm vlb}$ than expected.\\
\indent{}For $\beta_{\rm app}>0$, brightness temperatures have values in the range of 
\begin{equation}
    \gamma_{\rm j}-\sqrt{\gamma_{\rm j}^{2}-1}<\frac{T^{\rm a}_{\rm vlb,\nu^{\rm a}}}{T_{\rm int,\nu^{\rm e}}}<\gamma_{\rm j}+\sqrt{\gamma_{\rm j}^{2}-1}.
\end{equation}
Apparent speeds of approximately an order of magnitude greater than the current maximum observed $\beta_{\rm app}$ are required for the envelope to cover the high $T_{\rm vlb}$ outliers. Without observational evidence stating otherwise, we find it unlikely that an underestimated $\beta_{\rm app}$ can fully account for the outliers. Alternatively, we may consider horizontally shifting the envelope to the right to match the maximum $T_{\rm vlb}$ (i.e., increasing $T_{\rm int}$ by approximately an order of magnitude). In such a case, there would be a significant number of sources with a low $T_{\rm vlb}$ that lie above the envelope, with significantly fewer than 75\% of sources below the $\beta_{\rm max}$ line. This may be due to the insufficient completeness of our sample of sources, particularly toward the lower flux density limit. Based on these points, jointly with the arguments regarding unresolved cores, we consider the $T_{\rm int}$ determined from the population analysis to represent a lower limit on the true $T_{\rm int}$.\\
\indent{}Due to these considerations, we set upper and lower limits on $T_{\rm int}$.
These limits are given in Table~\ref{tab:freqdep_Tbint} and also plotted in Figure~\ref{fig:multifreqTbint}. It should be noted that \citet{2019A&A...622A..92N} utilizes an alternative population model \citep{2000A&A...364..391L} to find values of $\log_{10}T_{\rm int}\textrm{[K]}=11.32_{-0.02}^{+0.01}$, which is two orders of magnitude larger than those found in this paper and more comparable to the variability-based estimates. Accelerating jets in the millimeter core region \citep[e.g.,][]{2016ApJ...826..135L,2025A&A...695A.233R} and varying $T_{\rm int}$ with source variability \citep{2006ApJ...642L.115H} combined with a distribution of $\gamma_{\rm j}$ over multiple sources \citep[e.g.,][]{2021ApJ...923...67H} may introduce additional offsets between the different population studies (Appendix~\ref{sec:app_estimation_of_Tint}). Future experiments consisting of regular (e.g., less than monthly) multifrequency VLBI observations of a complete sample of sources may allow us to more concretely determine the potential frequency dependence of $T_{\rm int}$ as well as to resolve the offsets between different methods.\\
\begin{table}[t!]
\caption{\label{tab:freqdep_Tbint}Frequency dependent constraints on $T_{\rm int}$.}
\centering
\begin{tabular}{crcl}
\hline\hline
$\nu^{\rm o}$ &  \multicolumn{3}{c}{$\log_{10}T_{\rm int}$} \\
$(\textrm{GHz})$ & \multicolumn{3}{c}{$(\log_{10}\textrm{K})$} \\
\hline
& \multicolumn{3}{c}{$H_{0}=67.4\textrm{ km/s/Mpc}$}\\
\hline
15 & $10.13\tablefootmark{a}<$ & $\log_{10}T_{\rm int}$ & $<11.56$ \\
24 & $9.73<$ & $\log_{10}T_{\rm int}$ \\
43 & $9.12<$ & $\log_{10}T_{\rm int}$ & $<11.65$ \\
86 & $9.29<$ & $\log_{10}T_{\rm int}$ \\
\hline
\end{tabular}
\tablefoot{Constraints on $\log_{10}T_{\rm int}$ of AGN radio cores. Upper limits come from source variability (Section~\ref{sec:Tint_var}) and lower limits from a population analysis (Section~\ref{sec:Tint_pop}). For all values, a uniform disk was assumed for the core geometry. Each column corresponds to $\nu^{\rm o}$ (the observing frequency in GHz) and $\log_{10}T_{\rm int}$ (the lower and upper bounds on the logarithm of $T_{\rm int}$).
\tablefoottext{a}{Value from \citet{2014JKAS...47..303L}.}}
\end{table}

%

\section{Conclusions}
We have investigated the $T_{\rm int}$ of the cores of AGNs utilizing both flux variability and a population analysis of $T_{\rm vlb}$. During this process, we refined the equations in \citet{2023MNRAS.521L..44H} to properly account for the assumed source geometry, leading to a factor of approximately two lower value of $T_{\rm int}$ compared to the original equations. 
Assuming a uniform disk geometry, we placed upper and lower limits on $\log_{10}T_{\rm int}$ at various frequencies between 15 and 86~GHz. We were not able to constrain any frequency dependence on $T_{\rm int}$. Resolution of the offsets between the different methods as well as more detailed investigation of the redshift and frequency dependence of $T_{\rm int}$ may be possible with tailored, regular (high) cadence (quasi-)simultaneous multifrequency observations of a complete sample of sources spanning a wide redshift range.

\begin{acknowledgements}
We thank the anonymous reviewer for valuable comments and suggestions that helped to improve the paper.
This work was supported by the National Research Foundation of Korea (NRF) grant funded by the Korea government (MIST) (2020R1A2C2009003, RS-2025-00562700). This research has made use of data from the MOJAVE database that is maintained by the MOJAVE team \citep{2018ApJS..234...12L}
This study makes use of VLBA data from the VLBA-BU Blazar Monitoring Program (BEAM-ME and VLBA-BU-BLAZAR; http://www.bu.edu/blazars/BEAM-ME.html), funded by NASA through the Fermi Guest Investigator Program. The VLBA is an instrument of the National Radio Astronomy Observatory. The National Radio Astronomy Observatory is a facility of the National Science Foundation operated by Associated Universities, Inc.
\end{acknowledgements}

%

\bibliographystyle{bibtex/aa}
\bibliography{aa56473-25}

\begin{thebibliography}{40}
\expandafter\ifx\csname natexlab\endcsname\relax\def\natexlab#1{#1}\fi

\bibitem[{{Adame} {et~al.}(2025){Adame}, {Aguilar}, {Ahlen}, {Alam},
  {Alexander}, {Alvarez}, {Alves}, {Anand}, {Andrade}, {Armengaud}, {Avila},
  {Aviles}, {Awan}, {Bahr-Kalus}, {Bailey}, {Baltay}, {Bault}, {Behera},
  {BenZvi}, {Bera}, {Beutler}, {Bianchi}, {Blake}, {Blum}, {Brieden},
  {Brodzeller}, {Brooks}, {Buckley-Geer}, {Burtin}, {Calderon}, {Canning},
  {Carnero Rosell}, {Cereskaite}, {Cervantes-Cota}, {Chabanier}, {Chaussidon},
  {Chaves-Montero}, {Chen}, {Chen}, {Claybaugh}, {Cole}, {Cuceu}, {Davis},
  {Dawson}, {de la Macorra}, {de Mattia}, {Deiosso}, {Dey}, {Dey}, {Ding},
  {Doel}, {Edelstein}, {Eftekharzadeh}, {Eisenstein}, {Elliott}, {Fagrelius},
  {Fanning}, {Ferraro}, {Ereza}, {Findlay}, {Flaugher}, {Font-Ribera},
  {Forero-S{\'a}nchez}, {Forero-Romero}, {Frenk}, {Garcia-Quintero},
  {Gazta{\~n}aga}, {Gil-Mar{\'\i}n}, {Gontcho a Gontcho}, {Gonzalez-Morales},
  {Gonzalez-Perez}, {Gordon}, {Green}, {Gruen}, {Gsponer}, {Gutierrez}, {Guy},
  {Hadzhiyska}, {Hahn}, {Hanif}, {Herrera-Alcantar}, {Honscheid}, {Howlett},
  {Huterer}, {Ir{\v{s}}i{\v{c}}}, {Ishak}, {Juneau}, {Kara{\c{c}}ayl{\i}},
  {Kehoe}, {Kent}, {Kirkby}, {Kremin}, {Krolewski}, {Lai}, {Lan}, {Landriau},
  {Lang}, {Lasker}, {Le Goff}, {Le Guillou}, {Leauthaud}, {Levi}, {Li},
  {Linder}, {Lodha}, {Magneville}, {Manera}, {Margala}, {Martini}, {Maus},
  {McDonald}, {Medina-Varela}, {Meisner}, {Mena-Fern{\'a}ndez}, {Miquel},
  {Moon}, {Moore}, {Moustakas}, {Mueller}, {Mu{\~n}oz-Guti{\'e}rrez}, {Myers},
  {Nadathur}, {Napolitano}, {Neveux}, {Newman}, {Nguyen}, {Nie}, {Niz},
  {Noriega}, {Padmanabhan}, {Paillas}, {Palanque-Delabrouille}, {Pan},
  {Penmetsa}, {Percival}, {Pieri}, {Pinon}, {Poppett}, {Porredon}, {Prada},
  {P{\'e}rez-Fern{\'a}ndez}, {P{\'e}rez-R{\`a}fols}, {Rabinowitz}, {Raichoor},
  {Ram{\'\i}rez-P{\'e}rez}, {Ramirez-Solano}, {Rashkovetskyi}, {Ravoux},
  {Rezaie}, {Rich}, {Rocher}, {Rockosi}, {Roe}, {Rosado-Marin}, {Ross},
  {Rossi}, {Ruggeri}, {Ruhlmann-Kleider}, {Samushia}, {Sanchez}, {Saulder},
  {Schlafly}, {Schlegel}, {Schubnell}, {Seo}, {Shafieloo}, {Sharples},
  {Silber}, {Slosar}, {Smith}, {Sprayberry}, {Tan}, {Tarl{\'e}}, {Taylor},
  {Trusov}, {Ure{\~n}a-L{\'o}pez}, {Vaisakh}, {Valcin}, {Valdes},
  {Vargas-Maga{\~n}a}, {Verde}, {Walther}, {Wang}, {Wang}, {Weaver},
  {Weaverdyck}, {Wechsler}, {Weinberg}, {White}, {Yu}, {Yu}, {Yuan},
  {Y{\`e}che}, {Zaborowski}, {Zarrouk}, {Zhang}, {Zhao}, {Zhao}, {Zhou}, \&
  {Zhuang}}]{2025JCAP...02..021A}
{Adame}, A.~G., {Aguilar}, J., {Ahlen}, S., {et~al.} 2025, \jcap, 2025, 021

\bibitem[{{Boettcher} {et~al.}(2012){Boettcher}, {Harris}, \&
  {Krawczynski}}]{2012rjag.book.....B}
{Boettcher}, M., {Harris}, D.~E., \& {Krawczynski}, H. 2012, {Relativistic Jets
  from Active Galactic Nuclei}

\bibitem[{{Cheng} {et~al.}(2020){Cheng}, {An}, {Frey}, {Hong}, {He},
  {Kellermann}, {Lister}, {Lao}, {Li}, {Mohan}, {Yang}, {Wu}, {Zhang}, {Zhang},
  \& {Zhao}}]{2020ApJS..247...57C}
{Cheng}, X.~P., {An}, T., {Frey}, S., {et~al.} 2020, \apjs, 247, 57

\bibitem[{{Cheng} {et~al.}(2018){Cheng}, {An}, {Hong}, {Yang}, {Mohan},
  {Kellermann}, {Lister}, {Frey}, {Zhao}, {Zhang}, {Wu}, {Li}, \&
  {Zhang}}]{2018ApJS..234...17C}
{Cheng}, X.~P., {An}, T., {Hong}, X.~Y., {et~al.} 2018, \apjs, 234, 17

\bibitem[{{de Witt} {et~al.}(2023){de Witt}, {Jacobs}, {Gordon}, {Bietenholz},
  {Nickola}, \& {Bertarini}}]{2023AJ....165..139D}
{de Witt}, A., {Jacobs}, C.~S., {Gordon}, D., {et~al.} 2023, \aj, 165, 139

\bibitem[{D’Agostino \& Pearson(1973)}]{10.1093/biomet/60.3.613}
D’Agostino, R. \& Pearson, E.~S. 1973, Biometrika, 60, 613

\bibitem[{{Ghisellini} {et~al.}(1993){Ghisellini}, {Padovani}, {Celotti}, \&
  {Maraschi}}]{1993ApJ...407...65G}
{Ghisellini}, G., {Padovani}, P., {Celotti}, A., \& {Maraschi}, L. 1993, \apj,
  407, 65

\bibitem[{{Guijosa} \& {Daly}(1996)}]{1996ApJ...461..600G}
{Guijosa}, A. \& {Daly}, R.~A. 1996, \apj, 461, 600

\bibitem[{{Hodgson} {et~al.}(2020){Hodgson}, {L'Huillier}, {Liodakis}, {Lee},
  \& {Shafieloo}}]{2020MNRAS.495L..27H}
{Hodgson}, J.~A., {L'Huillier}, B., {Liodakis}, I., {Lee}, S.-S., \&
  {Shafieloo}, A. 2020, \mnras, 495, L27

\bibitem[{{Hodgson} {et~al.}(2023){Hodgson}, {L'Huillier}, {Liodakis}, {Lee},
  \& {Shafieloo}}]{2023MNRAS.521L..44H}
{Hodgson}, J.~A., {L'Huillier}, B., {Liodakis}, I., {Lee}, S.-S., \&
  {Shafieloo}, A. 2023, \mnras, 521, L44

\bibitem[{{Homan} {et~al.}(2021){Homan}, {Cohen}, {Hovatta}, {Kellermann},
  {Kovalev}, {Lister}, {Popkov}, {Pushkarev}, {Ros}, \&
  {Savolainen}}]{2021ApJ...923...67H}
{Homan}, D.~C., {Cohen}, M.~H., {Hovatta}, T., {et~al.} 2021, \apj, 923, 67

\bibitem[{{Homan} {et~al.}(2006){Homan}, {Kovalev}, {Lister}, {Ros},
  {Kellermann}, {Cohen}, {Vermeulen}, {Zensus}, \&
  {Kadler}}]{2006ApJ...642L.115H}
{Homan}, D.~C., {Kovalev}, Y.~Y., {Lister}, M.~L., {et~al.} 2006, \apjl, 642,
  L115

\bibitem[{{Hovatta} {et~al.}(2009){Hovatta}, {Valtaoja}, {Tornikoski}, \&
  {L{\"a}hteenm{\"a}ki}}]{2009A&A...494..527H}
{Hovatta}, T., {Valtaoja}, E., {Tornikoski}, M., \& {L{\"a}hteenm{\"a}ki}, A.
  2009, \aap, 494, 527

\bibitem[{{Hsu} {et~al.}(2023){Hsu}, {Koay}, {Matsushita}, {Hwang}, {Hovatta},
  {Kiehlmann}, {Readhead}, {Max-Moerbeck}, \& {Reeves}}]{2023MNRAS.525.5105H}
{Hsu}, P.-C., {Koay}, J.~Y., {Matsushita}, S., {et~al.} 2023, \mnras, 525, 5105

\bibitem[{{Jorstad} {et~al.}(2017){Jorstad}, {Marscher}, {Morozova},
  {Troitsky}, {Agudo}, {Casadio}, {Foord}, {G{\'o}mez}, {MacDonald}, {Molina},
  {L{\"a}hteenm{\"a}ki}, {Tammi}, \& {Tornikoski}}]{2017ApJ...846...98J}
{Jorstad}, S.~G., {Marscher}, A.~P., {Morozova}, D.~A., {et~al.} 2017, \apj,
  846, 98

\bibitem[{{Kellermann} \& {Pauliny-Toth}(1969)}]{1969ApJ...155L..71K}
{Kellermann}, K.~I. \& {Pauliny-Toth}, I.~I.~K. 1969, \apjl, 155, L71

\bibitem[{{Lee}(2013)}]{2013JKAS...46..243L}
{Lee}, S.-S. 2013, Journal of Korean Astronomical Society, 46, 243

\bibitem[{{Lee}(2014)}]{2014JKAS...47..303L}
{Lee}, S.-S. 2014, Journal of Korean Astronomical Society, 47, 303

\bibitem[{{Lee} {et~al.}(2008){Lee}, {Lobanov}, {Krichbaum}, {Witzel},
  {Zensus}, {Bremer}, {Greve}, \& {Grewing}}]{2008AJ....136..159L}
{Lee}, S.-S., {Lobanov}, A.~P., {Krichbaum}, T.~P., {et~al.} 2008, \aj, 136,
  159

\bibitem[{{Lee} {et~al.}(2016){Lee}, {Lobanov}, {Krichbaum}, \&
  {Zensus}}]{2016ApJ...826..135L}
{Lee}, S.-S., {Lobanov}, A.~P., {Krichbaum}, T.~P., \& {Zensus}, J.~A. 2016,
  \apj, 826, 135

\bibitem[{{Liodakis} {et~al.}(2018){Liodakis}, {Hovatta}, {Huppenkothen},
  {Kiehlmann}, {Max-Moerbeck}, \& {Readhead}}]{2018ApJ...866..137L}
{Liodakis}, I., {Hovatta}, T., {Huppenkothen}, D., {et~al.} 2018, \apj, 866,
  137

\bibitem[{{Liodakis} {et~al.}(2017){Liodakis}, {Marchili}, {Angelakis},
  {Fuhrmann}, {Nestoras}, {Myserlis}, {Karamanavis}, {Krichbaum}, {Sievers},
  {Ungerechts}, \& {Zensus}}]{2017MNRAS.466.4625L}
{Liodakis}, I., {Marchili}, N., {Angelakis}, E., {et~al.} 2017, \mnras, 466,
  4625

\bibitem[{{Lister} {et~al.}(2018){Lister}, {Aller}, {Aller}, {Hodge}, {Homan},
  {Kovalev}, {Pushkarev}, \& {Savolainen}}]{2018ApJS..234...12L}
{Lister}, M.~L., {Aller}, M.~F., {Aller}, H.~D., {et~al.} 2018, \apjs, 234, 12

\bibitem[{{Lister} {et~al.}(2019){Lister}, {Homan}, {Hovatta}, {Kellermann},
  {Kiehlmann}, {Kovalev}, {Max-Moerbeck}, {Pushkarev}, {Readhead}, {Ros}, \&
  {Savolainen}}]{2019ApJ...874...43L}
{Lister}, M.~L., {Homan}, D.~C., {Hovatta}, T., {et~al.} 2019, \apj, 874, 43

\bibitem[{{Lister} {et~al.}(2021){Lister}, {Homan}, {Kellermann}, {Kovalev},
  {Pushkarev}, {Ros}, \& {Savolainen}}]{2021ApJ...923...30L}
{Lister}, M.~L., {Homan}, D.~C., {Kellermann}, K.~I., {et~al.} 2021, \apj, 923,
  30

\bibitem[{{Lobanov}(2015)}]{2015A&A...574A..84L}
{Lobanov}, A. 2015, \aap, 574, A84

\bibitem[{{Lobanov} {et~al.}(2000){Lobanov}, {Krichbaum}, {Graham}, {Witzel},
  {Kraus}, {Zensus}, {Britzen}, {Greve}, \& {Grewing}}]{2000A&A...364..391L}
{Lobanov}, A.~P., {Krichbaum}, T.~P., {Graham}, D.~A., {et~al.} 2000, \aap,
  364, 391

\bibitem[{{Nair} {et~al.}(2019){Nair}, {Lobanov}, {Krichbaum}, {Ros}, {Zensus},
  {Kovalev}, {Lee}, {Mertens}, {Hagiwara}, {Bremer}, {Lindqvist}, \& {de
  Vicente}}]{2019A&A...622A..92N}
{Nair}, D.~G., {Lobanov}, A.~P., {Krichbaum}, T.~P., {et~al.} 2019, \aap, 622,
  A92

\bibitem[{{Pearson}(1995)}]{1995ASPC...82..267P}
{Pearson}, T.~J. 1995, in Astronomical Society of the Pacific Conference
  Series, Vol.~82, Very Long Baseline Interferometry and the VLBA, ed. J.~A.
  {Zensus}, P.~J. {Diamond}, \& P.~J. {Napier}, 267

\bibitem[{{Planck Collaboration} {et~al.}(2020){Planck Collaboration},
  {Aghanim}, {Akrami}, {Ashdown}, {Aumont}, {Baccigalupi}, {Ballardini},
  {Banday}, {Barreiro}, {Bartolo}, {Basak}, {Battye}, {Benabed}, {Bernard},
  {Bersanelli}, {Bielewicz}, {Bock}, {Bond}, {Borrill}, {Bouchet}, {Boulanger},
  {Bucher}, {Burigana}, {Butler}, {Calabrese}, {Cardoso}, {Carron},
  {Challinor}, {Chiang}, {Chluba}, {Colombo}, {Combet}, {Contreras}, {Crill},
  {Cuttaia}, {de Bernardis}, {de Zotti}, {Delabrouille}, {Delouis}, {Di
  Valentino}, {Diego}, {Dor{\'e}}, {Douspis}, {Ducout}, {Dupac}, {Dusini},
  {Efstathiou}, {Elsner}, {En{\ss}lin}, {Eriksen}, {Fantaye}, {Farhang},
  {Fergusson}, {Fernandez-Cobos}, {Finelli}, {Forastieri}, {Frailis},
  {Fraisse}, {Franceschi}, {Frolov}, {Galeotta}, {Galli}, {Ganga},
  {G{\'e}nova-Santos}, {Gerbino}, {Ghosh}, {Gonz{\'a}lez-Nuevo}, {G{\'o}rski},
  {Gratton}, {Gruppuso}, {Gudmundsson}, {Hamann}, {Handley}, {Hansen},
  {Herranz}, {Hildebrandt}, {Hivon}, {Huang}, {Jaffe}, {Jones}, {Karakci},
  {Keih{\"a}nen}, {Keskitalo}, {Kiiveri}, {Kim}, {Kisner}, {Knox},
  {Krachmalnicoff}, {Kunz}, {Kurki-Suonio}, {Lagache}, {Lamarre}, {Lasenby},
  {Lattanzi}, {Lawrence}, {Le Jeune}, {Lemos}, {Lesgourgues}, {Levrier},
  {Lewis}, {Liguori}, {Lilje}, {Lilley}, {Lindholm}, {L{\'o}pez-Caniego},
  {Lubin}, {Ma}, {Mac{\'\i}as-P{\'e}rez}, {Maggio}, {Maino}, {Mandolesi},
  {Mangilli}, {Marcos-Caballero}, {Maris}, {Martin}, {Martinelli},
  {Mart{\'\i}nez-Gonz{\'a}lez}, {Matarrese}, {Mauri}, {McEwen}, {Meinhold},
  {Melchiorri}, {Mennella}, {Migliaccio}, {Millea}, {Mitra},
  {Miville-Desch{\^e}nes}, {Molinari}, {Montier}, {Morgante}, {Moss}, {Natoli},
  {N{\o}rgaard-Nielsen}, {Pagano}, {Paoletti}, {Partridge}, {Patanchon},
  {Peiris}, {Perrotta}, {Pettorino}, {Piacentini}, {Polastri}, {Polenta},
  {Puget}, {Rachen}, {Reinecke}, {Remazeilles}, {Renzi}, {Rocha}, {Rosset},
  {Roudier}, {Rubi{\~n}o-Mart{\'\i}n}, {Ruiz-Granados}, {Salvati}, {Sandri},
  {Savelainen}, {Scott}, {Shellard}, {Sirignano}, {Sirri}, {Spencer},
  {Sunyaev}, {Suur-Uski}, {Tauber}, {Tavagnacco}, {Tenti}, {Toffolatti},
  {Tomasi}, {Trombetti}, {Valenziano}, {Valiviita}, {Van Tent}, {Vibert},
  {Vielva}, {Villa}, {Vittorio}, {Wandelt}, {Wehus}, {White}, {White},
  {Zacchei}, \& {Zonca}}]{2020A&A...641A...6P}
{Planck Collaboration}, {Aghanim}, N., {Akrami}, Y., {et~al.} 2020, \aap, 641,
  A6

\bibitem[{{Readhead}(1994)}]{1994ApJ...426...51R}
{Readhead}, A. C.~S. 1994, \apj, 426, 51

\bibitem[{{Riess} {et~al.}(2022){Riess}, {Yuan}, {Macri}, {Scolnic}, {Brout},
  {Casertano}, {Jones}, {Murakami}, {Anand}, {Breuval}, {Brink}, {Filippenko},
  {Hoffmann}, {Jha}, {D'arcy Kenworthy}, {Mackenty}, {Stahl}, \&
  {Zheng}}]{2022ApJ...934L...7R}
{Riess}, A.~G., {Yuan}, W., {Macri}, L.~M., {et~al.} 2022, \apjl, 934, L7

\bibitem[{{R{\"o}der} {et~al.}(2025){R{\"o}der}, {Wielgus}, {Lobanov},
  {Krichbaum}, {Nair}, {Lee}, {Ros}, {Fish}, {Blackburn}, {Chan}, {Issaoun},
  {Janssen}, {Johnson}, {Doeleman}, {Bower}, {Crew}, {Tilanus}, {Savolainen},
  {Violette Impellizzeri}, {Alberdi}, {Baczko}, {G{\'o}mez}, {Lu}, {Paraschos},
  {Traianou}, {Goddi}, {Kim}, {Lisakov}, {Kovalev}, {Voitsik}, {Sokolovsky},
  {Akiyama}, {Albentosa-Ru{\'\i}z}, {Alef}, {Algaba}, {Anantua}, {Asada},
  {Azulay}, {Bach}, {Ball}, {Balokovi{\'c}}, {Bandyopadhyay}, {Barrett},
  {Baub{\"o}ck}, {Benson}, {Bintley}, {Blundell}, {Bouman}, {Bremer},
  {Brinkerink}, {Brissenden}, {Britzen}, {Broderick}, {Broguiere}, {Bronzwaer},
  {Bustamante}, {Byun}, {Carlstrom}, {Ceccobello}, {Chael}, {Chang},
  {Chatterjee}, {Chatterjee}, {Chen}, {Chen}, {Cheng}, {Cho}, {Christian},
  {Conroy}, {Conway}, {Cordes}, {Crawford}, {Cruz-Osorio}, {Cui}, {Curd},
  {Dahale}, {Davelaar}, {De Laurentis}, {Deane}, {Dempsey}, {Desvignes},
  {Dexter}, {Dhruv}, {Dihingia}, {Dougall}, {Dzib}, {Eatough}, {Emami},
  {Falcke}, {Farah}, {Fomalont}, {Alyson Ford}, {Foschi}, {Fraga-Encinas},
  {Freeman}, {Friberg}, {Fromm}, {Fuentes}, {Galison}, {Gammie}, {Garc{\'\i}a},
  {Gentaz}, {Georgiev}, {Gold}, {G{\'o}mez-Ruiz}, {Gu}, {Gurwell}, {Hada},
  {Haggard}, {Haworth}, {Hecht}, {Hesper}, {Heumann}, {Ho}, {Ho}, {Honma},
  {Huang}, {Huang}, {Hughes}, {Ikeda}, {Inoue}, {James}, {Jannuzi}, {Jeter},
  {Jiang}, {Jim{\'e}nez-Rosales}, {Jorstad}, {Joshi}, {Jung}, {Karami},
  {Karuppusamy}, {Kawashima}, {Keating}, {Kettenis}, {Kim}, {Kim}, {Kim},
  {Kim}, {Kino}, {Yi Koay}, {Kocherlakota}, {Kofuji}, {Koyama}, {Kramer},
  {Kramer}, {Kramer}, {Kuo}, {La Bella}, {Lauer}, {Lee}, {Leung}, {Levis},
  {Li}, {Lico}, {Lindahl}, {Lindqvist}, {Liu}, {Liu}, {Liuzzo}, {Lo},
  {Loinard}, {Lonsdale}, {Lowitz}, {MacDonald}, {Mao}, {Marchili}, {Markoff},
  {Marrone}, {Marscher}, {Mart{\'\i}-Vidal}, {Matsushita}, {Matthews},
  {Medeiros}, {Menten}, {Michalik}, {Mizuno}, {Mizuno}, {Moran}, {Moriyama},
  {Moscibrodzka}, {Mulaudzi}, {M{\"u}ller}, {M{\"u}ller}, {Mus}, {Musoke},
  {Myserlis}, {Nadolski}, {Nagai}, {Nagar}, {Nakamura}, {Narayanan},
  {Natarajan}, {Nathanail}, {Navarro Fuentes}, {Neilsen}, {Neri}, {Ni},
  {Noutsos}, {Nowak}, \& {Oh}}]{2025A&A...695A.233R}
{R{\"o}der}, J., {Wielgus}, M., {Lobanov}, A.~P., {et~al.} 2025, \aap, 695,
  A233

\bibitem[{{Savolainen} {et~al.}(2006){Savolainen}, {Wiik}, {Valtaoja}, \&
  {Tornikoski}}]{2006A&A...446...71S}
{Savolainen}, T., {Wiik}, K., {Valtaoja}, E., \& {Tornikoski}, M. 2006, \aap,
  446, 71

\bibitem[{Shapiro \& Wilk(1965)}]{eb32428d-e089-3d0c-8541-5f3e8f273532}
Shapiro, S.~S. \& Wilk, M.~B. 1965, Biometrika, 52, 591

\bibitem[{{Strauss} {et~al.}(1992){Strauss}, {Huchra}, {Davis}, {Yahil},
  {Fisher}, \& {Tonry}}]{1992ApJS...83...29S}
{Strauss}, M.~A., {Huchra}, J.~P., {Davis}, M., {et~al.} 1992, \apjs, 83, 29

\bibitem[{Thompson {et~al.}(2017)Thompson, Moran, \& Swenson}]{Thompson2017}
Thompson, A.~R., Moran, J.~M., \& Swenson, G.~W. 2017, Calibration and Imaging
  (Cham: Springer International Publishing), 485--549

\bibitem[{{Tingay} {et~al.}(2001){Tingay}, {Preston}, {Lister}, {Piner},
  {Murphy}, {Jones}, {Meier}, {Pearson}, {Readhead}, {Hirabayashi}, {Murata},
  {Kobayashi}, \& {Inoue}}]{2001ApJ...549L..55T}
{Tingay}, S.~J., {Preston}, R.~A., {Lister}, M.~L., {et~al.} 2001, \apjl, 549,
  L55

\bibitem[{{Weaver} {et~al.}(2022){Weaver}, {Jorstad}, {Marscher}, {Morozova},
  {Troitsky}, {Agudo}, {G{\'o}mez}, {L{\"a}hteenm{\"a}ki}, {Tammi}, \&
  {Tornikoski}}]{2022ApJS..260...12W}
{Weaver}, Z.~R., {Jorstad}, S.~G., {Marscher}, A.~P., {et~al.} 2022, \apjs,
  260, 12

\bibitem[{{Wiik} \& {Valtaoja}(2001)}]{2001A&A...366.1061W}
{Wiik}, K. \& {Valtaoja}, E. 2001, \aap, 366, 1061

\end{thebibliography}







   
  



\begin{appendix}




\section{Deriving the causality-constrained angular diameter distance}\label{sec:app_eqderv}
\subsection{Intensity, total flux density, and brightness temperature}\label{sec:totfluxnTb}
The intensity of a source for the different geometries considered may be parameterized as follows \citep[e.g.,][]{Thompson2017}:\\
For a uniform disk with radius $a$,
\begin{equation}
    \frac{I_{\nu}\left(r\right)}{I_{\nu,0}}=\Pi\left(\frac{r}{a}\right).
    \label{eq:disk_intensity}
\end{equation}
For a circular Gaussian with a linear FWHM $l_{\rm FWHM}=\sqrt{8\ln{2}}a$,
\begin{equation}
    \frac{I_{\nu}\left(r\right)}{I_{\nu,0}} = \exp{\left(-\frac{r^2}{2a^2}\right)}. 
    \label{eq:gauss_intensity}
\end{equation}
Finally, for an optically thin uniform sphere with radius $a$,
\begin{equation}
    \frac{I_{\nu}\left(r\right)}{I_{\nu,0}} = \sqrt{1-\left[\frac{r}{a}\right]^2}\Pi\left(\frac{r}{a}\right).
    \label{eq:sphere_intensity}
\end{equation}
In the equations, $r$ is the radial offset from the center, and $I_{\nu,0}$ is the intensity at $r=0$.
The function $\Pi\left(x\right)$ is defined such that
\begin{equation}
    \Pi\left(x\right)=\left\{\begin{array}{l}
1\textrm{ : }-1\leq x\leq1 \\\\
0\textrm{ : else} \end{array}\right. .
\end{equation}
To find the total flux density, $S_{\nu}$, that would be observed, we integrated over the full intensity distribution:
\begin{equation}
    S_{\nu}=\int{I_{\nu}d\Omega}.
\end{equation}
Starting with a circular Gaussian, we have
\begin{equation}
    S_{\rm Gauss, \nu}
    =\int{I_{\nu,0}\exp{\left(-\frac{r^2}{2a^2}\right)}d\Omega}
    =\frac{\pi I_{\nu,0}l_{\rm FWHM}^2}{4\ln{2}D_{\rm A}^2}
    =\frac{\pi I_{\nu,0}}{4\ln{2}}\theta_{\rm FWHM}^2,
\end{equation}
where we used $l_{\rm FWHM}=\sqrt{8\ln{2}}a$ and $\theta_{\rm FWHM}=l_{\rm FWHM}/D_{\rm A}$. For a uniform disk, we have
\begin{equation}
    S_{\rm disk, \nu}
    =\int{I_{\nu,0}\Pi\left(\frac{r}{a}\right)d\Omega}
    =\frac{\pi I_{\nu,0}a^2}{D_{\rm A}^2}
    =\pi I_{\nu,0}\theta_{\rm a}^2,
\end{equation}
where $\theta_{\rm a}=a/D_{\rm A}$. For a uniform sphere,
\begin{equation}
        S_{\rm sphere,\nu}
        =\int{I_{\nu,0}\sqrt{1-\left[\frac{r}{a}\right]^2}\Pi\left(\frac{r}{a}\right)d\Omega}
        =\frac{2\pi I_{\nu,0}a^2}{3D_{\rm A}^2}
        =\frac{2\pi I_{\nu,0}}{3}\theta_{\rm a}^2.
\end{equation}

\indent{}The intensity of a source is commonly referenced to an equivalent blackbody temperature, i.e., the brightness temperature, $T_{\rm b}$. In the Rayleigh-Jeans limit $\left(h\nu\ll k_{\rm B}T_{\rm b}\right)$, $I_{\nu}\approx2k_{\rm B}\nu^2T_{\rm b,\nu}/c^2$. Therefore, we have
\begin{equation}
    T_{\rm b,\nu, disk}=\frac{c^2}{2\pi k_{\rm B}\nu^2}\frac{D_{\rm A}^2}{a^2}S_{\nu}
    =\frac{c^2}{2\pi k_{\rm B}\nu^2}\frac{S_{\nu}}{\theta_{\rm a}^2}
    \label{eq:disk_tb_orig}
\end{equation}
for a uniform disk with a radius of $a$,
\begin{equation}
    T_{\rm b,\nu,Gauss}
    =\frac{2\ln{2}c^2}{\pi k_{\rm B}\nu^2}\frac{D_{\rm A}^2}{l_{\rm FWHM}^2}S_{\nu}
    =\frac{2\ln{2}c^2}{\pi k_{\rm B}\nu^2}\frac{S_{\nu}}{\theta_{\rm FWHM}^2}
    \label{eq:gauss_tb_orig_re}
\end{equation}
for a circular Gaussian with a FWHM of $l_{\rm FWHM}$, and
\begin{equation}
    T_{\rm b,\nu,sphere}
    =\frac{3c^2}{4\pi k_{\rm B}\nu^2}\frac{D_{\rm A}^2}{a^2}S_{\nu}
    =\frac{3c^2}{4\pi k_{\rm B}\nu^2}\frac{S_{\nu}}{\theta_{\rm a}^2}
    \label{eq:sphere_tb_orig}
\end{equation}
for a uniform sphere with a radius of $a$.
The angular size ($\theta_{\rm a}$, $\theta_{\rm FWHM}$) of the source can be measured with VLBI. We denote such VLBI-measured brightness temperatures as $T_{\rm vlb,\nu}$.\\
$\textrm{}$\indent{}We can also employ $a = c\Delta t$ to define a ``variability brightness temperature,'' $T_{\rm var,\nu}$. For a disk, this is 
\begin{equation}
    T_{\rm var,\nu, disk}=\frac{D_{\rm A}^2}{2\pi k_{\rm B}\nu^2\Delta t_{\nu}^2}S_{\nu},
\end{equation}
and for a sphere,
\begin{equation}
    T_{\rm var,\nu, sphere}=\frac{3D_{\rm A}^2}{4\pi k_{\rm B}\nu^2\Delta t_{\nu}^2}S_{\nu}.
\end{equation}
The total physical size of a Gaussian is not as well-defined and we do not consider the variability brightness temperature of a Gaussian source here. \\
$\textrm{}$\indent{} We briefly consider the measurement of $\theta_{\rm a}$ with VLBI. Modeling of VLBI data typically occurs in the ``visibility plane'', i.e., the Fourier space of the image plane and the actual measurements made with VLBI. The visibilties corresponding to Equations~(\ref{eq:disk_intensity}),~(\ref{eq:gauss_intensity}),~(\ref{eq:sphere_intensity}) are as follows\citep[e.g.,][]{Thompson2017}:\\
For a uniform disk with radius $a$,
\begin{equation}
    \frac{V\left(q\right)}{I_0}=\pi a^2\frac{J_1\left(2\pi aq\right)}{\pi aq}.
    \label{eq:disk_vis}
\end{equation}
For a circular Gaussian with linear FWHM $l_{\rm FWHM}=\sqrt{8\ln{2}}a$,
\begin{equation}
    \frac{V\left(q\right)}{I_0}=2\pi a^2e^{-2\pi^2a^2q^2}.
    \label{eq:gauss_vis}
\end{equation}
Finally, for an optically thin uniform sphere with radius $a$,
\begin{equation}
    \frac{V\left(q\right)}{I_0}=\sqrt{\frac{\pi}{2}}\left(2\pi a^2\right)\frac{J_{3/2}\left(2\pi aq\right)}{\left(2\pi aq\right)^{3/2}}.
    \label{eq:sphere_vis}
\end{equation}
In the equations, $J_{\rm n}$ is the Bessel function of the first kind. For the same value of $V\left(0\right)$, a Gaussian source with FWHM $l_{\rm FWHM}$, a disk with diameter $2a\approx1.6l_{\rm FWHM}$, and a sphere with diameter $2a\approx1.8l_{\rm FWHM}$ all have the same visibility FWHM \citep[e.g.,][]{1995ASPC...82..267P}. Without measurements at sufficiently long baselines such that we are able to observe the null points, and/or without sufficient signal-to-noise ratio, it may prove difficult to directly differentiate between the three geometries. Therefore, in the absence of sufficient data, we may consider model-fitting a Gaussian to the visibilities. Then, we may take the model-fit component to represent a Gaussian with FWHM $l_{\rm FWHM}$, a disk with radius $a\approx0.8l_{\rm FWHM}$, or a sphere with radius $a\approx0.9l_{\rm FWHM}$. Indeed, such an approximation was made in \citet{2020MNRAS.495L..27H,2023MNRAS.521L..44H} as well. \\
$\textrm{}$\indent{}We could then determine approximations to Equations~(\ref{eq:disk_tb_orig}),~(\ref{eq:sphere_tb_orig}) as follows:\\
For a uniform disk with radius $a$,
\begin{equation}
    T_{\rm b,\nu,disk}
    \approx \frac{c^2}{2\pi k_{\rm B}\nu^2}\frac{S_{\nu}}{\left(0.8\theta_{\rm FWHM}\right)^2}
    ,
\end{equation}
and for an optically thin uniform sphere with radius $a$,
\begin{equation}
    T_{\rm b,\nu,sphere}
    \approx\frac{3c^2}{4\pi k_{\rm B}\nu^2}\frac{S_{\nu}}{\left(0.9\theta_{\rm FWHM}\right)^2}
    .
\end{equation}
For the same value of $\theta_{\rm FWHM}$, $T_{\rm b,\nu,disk}\approx0.56T_{\rm b,\nu,Gauss}$, and $T_{\rm b,\nu,sphere}\approx0.67T_{\rm b,\nu,Gauss}$ \citep[e.g.,][]{2001ApJ...549L..55T, 2006A&A...446...71S}.
\subsection{Relativistic corrections to $T_{\rm b,\nu}$}
In total we can consider three frames of reference. We have the observer frame $\mathcal{F}^{\rm o}$ at $z_{\rm cos}=0$, in which the measurements are made. We then have the host-galaxy frame $\mathcal{F}^{\rm a}$ of the AGN at redshift $z_{\rm cos}$. Finally we have the jet emission frame $\mathcal{F}^{\rm e}$, moving at a speed $\beta$ relative to $\mathcal{F}^{\rm a}$ according to the bulk jet speed. First we discuss transformations between $\mathcal{F}^{\rm a}$ and $\mathcal{F}^{\rm e}$. Transformations between $\mathcal{F}^{\rm a}$ and $\mathcal{F}^{\rm e}$ are governed by special relativity. For $T_{\rm vlb,\nu}$ we have \citep[e.g.,][]{2012rjag.book.....B}
\begin{equation}
    T^{\rm a}_{\rm vlb,\nu^{\rm a}}
    =\left\{\begin{array}{l}
        \delta^{3+\alpha}T^{\rm e}_{\rm vlb,\nu^{\rm a}}\\\\
        \delta T^{\rm e}_{\rm vlb,\nu^{\rm e}}
    \end{array}\right.,
\end{equation}
where we assumed $S_{\nu}\propto\nu^{-\alpha}$. For $T_{\rm var,\nu}$ we have
\begin{equation}
    T^{\rm a}_{\rm var,\nu^{\rm a}}
    =\left\{\begin{array}{l}
        \delta^{5+\alpha-2\eta}T^{\rm e}_{\rm var,\nu^{\rm a}}
        \\\\
        \delta^3T^{\rm e}_{\rm var,\nu^{\rm e}}
    \end{array}\right.,
\end{equation}
where we assumed $S_{\nu}\propto\nu^{-\alpha}$ and $\Delta t_{\nu}\propto\nu^{-\eta}$. Regardless of spectral shape (i.e., $\alpha$) and frequency dependence of $\Delta t$ (i.e., $\eta$), we have
\begin{equation}
    T^{\rm a}_{\rm vlb,\nu^{\rm a}}=\delta T^{\rm e}_{\rm vlb,\nu^{\rm e}}
    \textrm{,}~T^{\rm a}_{\rm var,\nu^{\rm a}}=\delta^3T^{\rm e}_{\rm var,\nu^{\rm e}}.\label{eg:Tb_a_to_e}
\end{equation}
If we assume $T^{\rm e}_{\rm vlb,\nu^{\rm e}}=T^{\rm e}_{\rm var,\nu^{\rm e}}=T_{\rm int,\nu^{\rm e}}$ where $T_{\rm int,\nu^{\rm e}}$ is some intrinsic brightness temperature of the source, we have $T^{\rm a}_{\rm vlb,\nu^{\rm a}}=\delta T_{\rm int,\nu^{\rm e}}$ and $T^{\rm a}_{\rm var,\nu^{\rm a}}=\delta^3T_{\rm int,\nu^{\rm e}}$.
It should be noted that these equations hold regardless of whether the source is a disk, Gaussian, or sphere.\\
\indent The observables undergo further changes due to the cosmological evolution of the universe (transformations between $\mathcal{F}^{\rm o}$ and $\mathcal{F}^{\rm a}$). Relevant to the discussion in this paper, we have
    $\nu^{\rm a}/\nu^{\rm o}
    =\Delta t^{\rm o}/\Delta t^{\rm a}
    =S^{\rm a}_{\nu^{\rm a}}/S^{\rm o}_{\nu^{\rm o}}
    =\left(1+z_{\rm cos}\right)$.
Due to such cosmological corrections, we have 
\begin{equation}
\begin{split}
    T^{\rm a}_{\rm vlb,\nu^{\rm a}}
    &=\left(1+z_{\rm cos}\right)T^{\rm o}_{\rm vlb,\nu^{\rm o}}\\
    T^{\rm a}_{\rm var,\nu^{\rm a}}
    &=\left(1+z_{\rm cos}\right)^{3}T^{\rm o}_{\rm var,\nu^{\rm o}}.
\end{split}
\end{equation}
Combined with Equation~(\ref{eg:Tb_a_to_e}), this leads to 
\begin{equation}
\begin{split}
    T^{\rm e}_{\rm vlb,\nu^{\rm e}}
    &=\frac{\left(1+z_{\rm cos}\right)}{\delta}T^{\rm o}_{\rm vlb,\nu^{\rm o}}\\
    T^{\rm e}_{\rm var,\nu^{\rm e}}
    &=\left(\frac{1+z_{\rm cos}}{\delta}\right)^{3}T^{\rm o}_{\rm var,\nu^{\rm o}}.
\end{split}
\end{equation}
\subsection{Variability angular diameter distance}
The variability angular diameter distance may now be found as
\begin{equation}
    D_{\rm A}
    =\frac{c\Delta t^{\rm e}_{\nu^{\rm e}}}{\theta_{\rm a}}
    =\frac{c\delta\Delta t^{\rm a}_{\nu^{\rm a}}}{\theta_{\rm a}}
    =\frac{c\delta\Delta t^{\rm o}_{\nu^{\rm o}}}{\left(1+z_{\rm cos}\right)\theta_{\rm a}}
    \label{eq:da_eq_wyc_v1},
\end{equation} 
as in \citet{2020MNRAS.495L..27H}. Following \citet{2023MNRAS.521L..44H}, if we wish to instead use $T_{\rm int}$ to cancel out $\delta$, we have
\begin{equation}
\begin{split}
    D_{\rm A}
    =\frac{c\delta\Delta t^{\rm a}_{\nu^{\rm a}}}{\theta_{\rm a}}
    =\frac{c\Delta t^{\rm a}_{\nu^{\rm a}}}{\theta_{\rm a}}\frac{T^{\rm a}_{\rm vlb,\nu^{\rm a}}}{T_{\rm int,\nu^{\rm e}}}
    =\frac{c\Delta t^{\rm o}_{\nu^{\rm o}}}{\theta_{\rm a}}\frac{T^{\rm o}_{\rm vlb,\nu^{\rm o}}}{T_{\rm int,\nu^{\rm e}}}.
    \label{eq:da_eq_wyc_v2_0}
\end{split}
\end{equation} 
For a disk or sphere, this reduces to
\begin{equation}
    D_{\rm A}
    =\left\{
    \begin{array}{ll}
        \frac{c^{3}\Delta t^{\rm o}_{\nu^{\rm o}}S^{\rm o}_{\nu^{\rm o}}}{2\pi k_{\rm B}T_{\rm int,\nu^{\rm e}}\nu^{\textrm{o}2}\theta_{\rm a}^3}& (\textrm{disk},\theta_{\rm a}\approx0.8\theta_{\rm FWHM}) 
        \\~\\
        \frac{3c^{3}\Delta t^{\rm o}_{\nu^{\rm o}}S^{\rm o}_{\nu^{\rm o}}}{4\pi k_{\rm B}T_{\rm int,\nu^{\rm e}}\nu^{\textrm{o}2}\theta_{\rm a}^3} & (\textrm{sphere},\theta_{\rm a}\approx0.9\theta_{\rm FWHM})
    \end{array}
    \right. .
\end{equation}
If we wish to rearrange the equations to assimilate the function of $T_{\rm b,Gauss}$, we have
\begin{equation}
    D_{\rm A}
    =X\frac{2\ln{2}c^3\Delta t^{\rm o}_{\nu^{\rm o}}S^{\rm o}_{\nu^{\rm o}}}{\pi k_{\rm B}T_{\rm int,\nu^{\rm e}}\nu^{\textrm{o}2}\theta_{\rm FWHM}^3},
    \label{eq:wyc_eq_nocos}
\end{equation}
where $X\approx0.70$ for a disk and $X\approx0.74$ for a sphere. Here $\theta_{\rm FWHM}$ is the FWHM of the Gaussian function used to model the visibilities of the source. Equation~(\ref{eq:wyc_eq_nocos}) may also be used to calculate the intrinsic brightness temperature by rearranging $T_{\rm int,\nu^{\rm e}}$ to the left hand side and $D_{\rm A}$ to the right hand side of the equation.\\
\indent{}It should be noted that our equations slightly differ from \citet{2023MNRAS.521L..44H}, who suggest that 
\begin{equation}
    D_{\rm A}\approx\frac{2\ln{2}c^3\Delta t^{\rm o}_{\nu^{\rm o}}S^{\rm o}_{\nu^{\rm o}}}{\pi k_{\rm B}T_{\rm int,\nu^{\rm e}}\nu^{\textrm{o}2}\theta_{\rm a}^3}
    \label{eq:hod2023_orig_v2},
\end{equation}
where $\theta_{\rm a}=0.8\theta_{\rm FWHM}$ for a disk and $\theta_{\rm a}=0.9\theta_{\rm FWHM}$ for a sphere (i.e., $X\approx1.95$ for a disk and $X\approx1.37$ for a sphere). The difference in the ``geometrical correction factors'' prescribed in Equation~(\ref{eq:hod2023_orig_v2}) from those derived in Equation~(\ref{eq:wyc_eq_nocos}) stems from the function of $T_{\rm vlb}$ used when canceling $\delta$ in Equation~(\ref{eq:da_eq_wyc_v1}). Equation~(\ref{eq:hod2023_orig_v2}) corrects for the measured angular size (i.e., from $\theta_{\rm FWHM}$ to $\theta_{\rm a}$) while approximating $T_{\rm vlb,\nu}$ as $T_{\rm b,\nu,Gauss}$. Equation~(\ref{eq:wyc_eq_nocos}) accounts for the appropriate analytical functions of $T_{\rm vlb,\nu}$ for the assumed geometry (i.e., Equations~(\ref{eq:disk_tb_orig}),~(\ref{eq:sphere_tb_orig}) respectively for a disk and spherical emission region) as well. We find that compared to Equation~(\ref{eq:wyc_eq_nocos}), the use of Equation~(\ref{eq:hod2023_orig_v2}) overestimates $D_{\rm A}$ by approximately a factor of 2.77 for a disk and 1.85 for a sphere, which, if left unaccounted for, results in a large systematic offset in $D_{\rm A}$ (or equivalently, $T_{\rm int}$).
\section{Biases in the estimation of $T_{\rm int}$}\label{sec:app_estimation_of_Tint}
\begin{figure}[t!]
    \centering
    \begin{subfigure}[t]{0.4\textwidth}
        \centering
        \includegraphics[width=\textwidth]{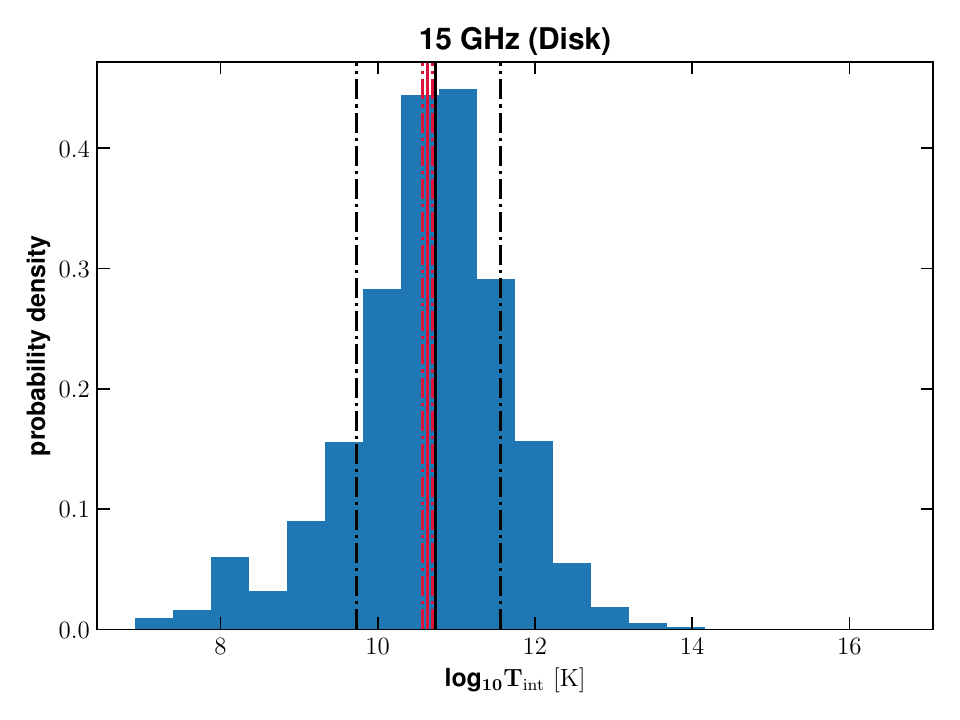}
        \caption{Combined probability density of the 15~GHz sample}
    \end{subfigure}
    \\
    \begin{subfigure}[t]{0.4\textwidth}
        \centering
        \includegraphics[width=\textwidth]{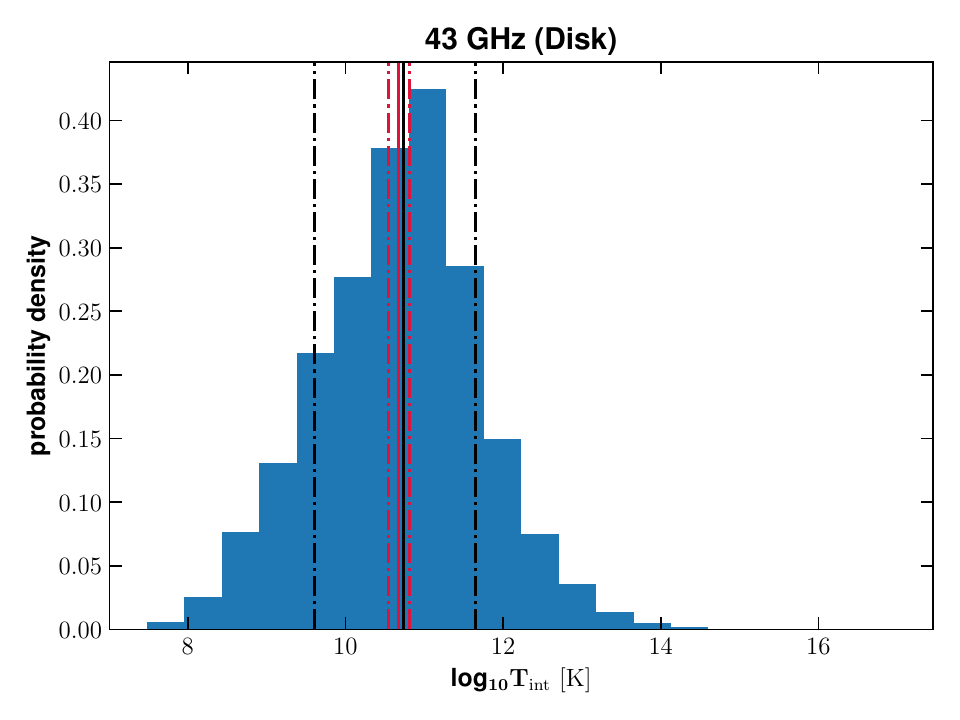}
        \caption{Combined probability density of the 43~GHz sample}
    \end{subfigure}
    \caption{Estimated $T_{\rm int}$ at 15~GHz (upper row) and at 43~GHz (lower row). }
    \label{fig:Tint_var_per_source_hist}
\end{figure}

\begin{figure*}[t!]
    \centering
    \begin{subfigure}[t]{0.32\textwidth}
        \centering
        \includegraphics[width=\textwidth]{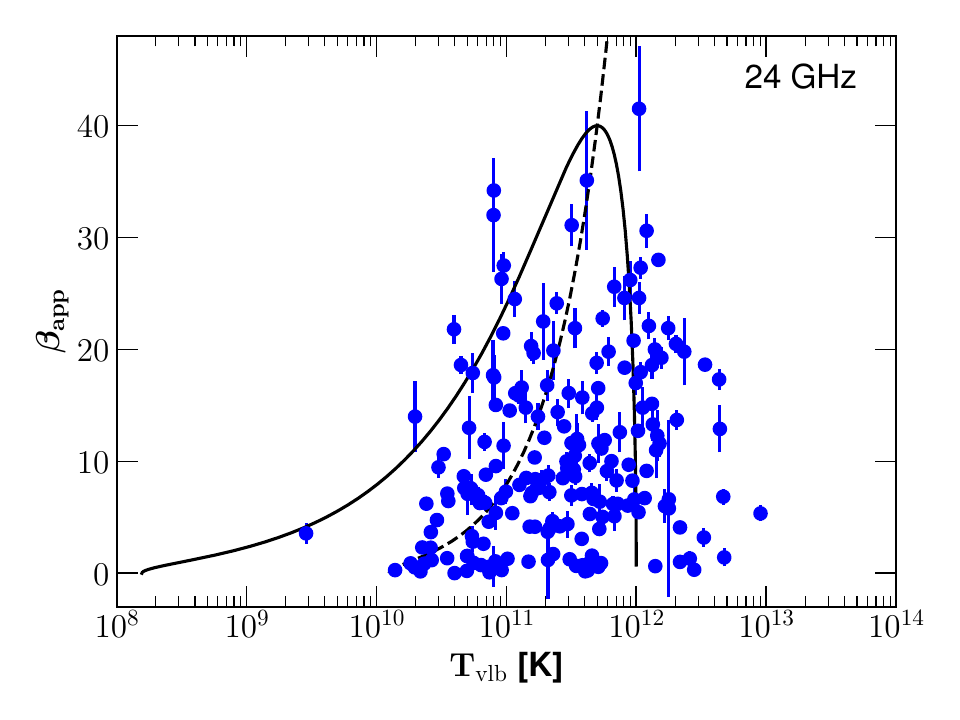}
    \end{subfigure}
    ~ 
    \begin{subfigure}[t]{0.32\textwidth}
        \centering
        \includegraphics[width=\textwidth]{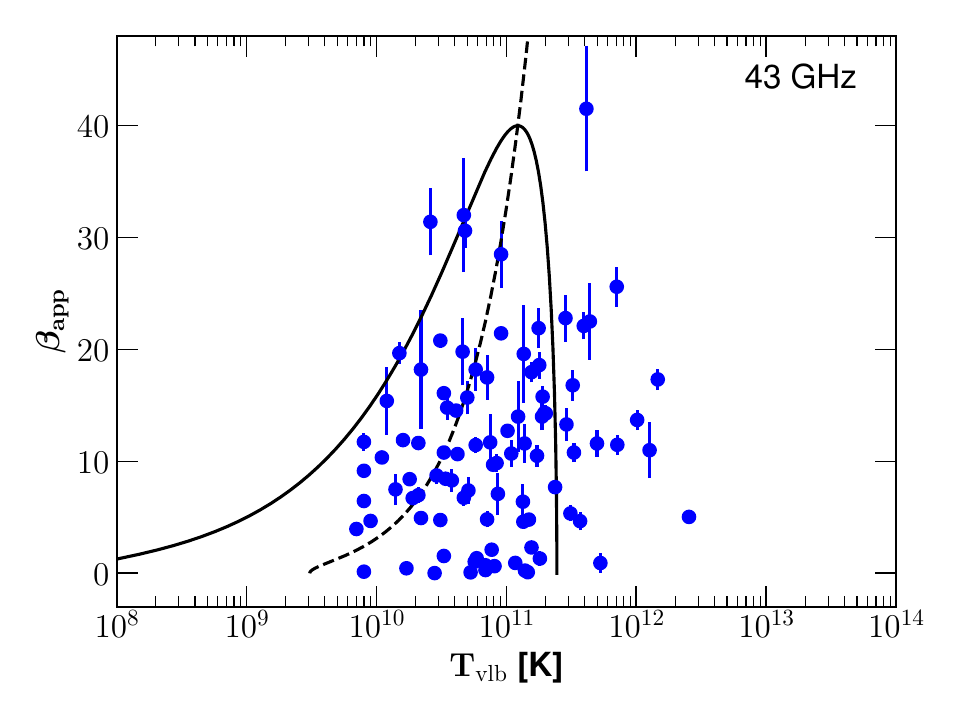}
    \end{subfigure}
    ~
    \begin{subfigure}[t]{0.32\textwidth}
        \centering
        \includegraphics[width=\textwidth]{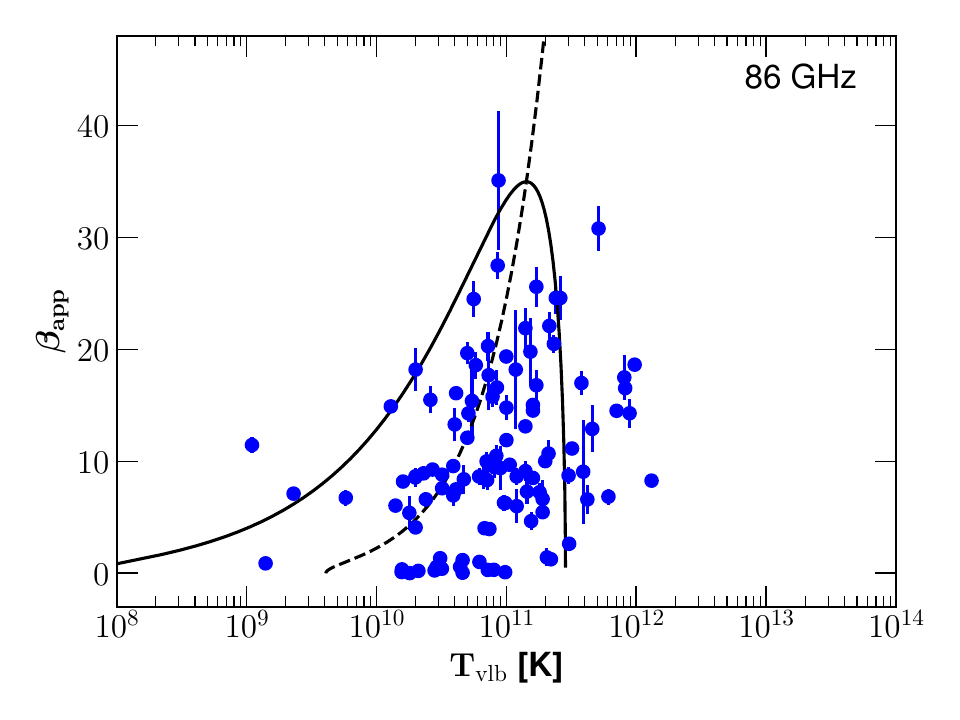}
    \end{subfigure}
    \caption{Estimated $T_{\rm int}$ at 24~GHz (left), 43~GHz (center), and 86~GHz (right) using a population study (see Section~\ref{sec:app_estimation_of_Tint} for details). }
    \label{fig:Tint_pop_per_source}
\end{figure*}

In the main text, we describe the expectation of a log-normal distribution, and no trend in $T_{\rm int}$ as a function of cadence. The two-point estimate of $\tau_{i+1,i}$ relies on significant flux variability measurements with a cadence comparable to or better than the variability timescale of the flare in question. Estimates of $\tau_{i+1,i}$ may be biased toward larger values, should the conditions not be met. As a simple test, we alternate the constraint on $t_{i+1}-t_{i}$ to be less than 30~days, 50~days, 100~days, 500~days, and unconstrained. Values of $\log_{10}T_{\rm int}\textrm{~[K]}$, as well as the sample average of $\log_{10}T_{\rm int}\textrm{~[K]}$ (i.e., $\left<\log_{10}T_{\rm int}\textrm{~[K]}\right>$), are given in Table~\ref{tab:tint_var_bias}. Histograms of the estimated $\log_{10}T_{\rm int}$~[K] with $t_{i+1}-t_{i}<30$~days may be found in Figure~\ref{fig:Tint_var_per_source_hist}. The black solid and dot-dashed vertical lines are the median and corresponding one-sigma confidence interval on $\log_{10}T_{\rm int}$~[K]. The red solid and dot-dashed horizontal lines are the same, but for $\left<\log_{10}T_{\rm int}\textrm{~[K]}\right>$.
Estimates of both $\log_{10}T_{\rm int}\textrm{~[K]}$ and $\left<\log_{10}T_{\rm int}\textrm{~[K]}\right>$ tend to be higher when the constraint on $t_{i+1}-t_{i}$ changes from 30~days to unconstrained.\\
\indent{}It should be noted that the number of individual sources that remain within each filter varies greatly (from 75 to 356 sources) for the 15~GHz data. This is primarily due to the varying (and somewhat lower) cadence of the MOJAVE observations. The number of sources does not change as much (37-38) at 43~GHz, due to the regular (bi-)monthly observations of the Boston University monitoring program. When repeating the analysis at 15~GHz with the source sample limited to the 75 sources that pass all 5 constraint sets, we find that the data-constraint-induced biases still exist (Table~\ref{tab:tint_var_bias_7src}). Values of $\log_{10}T_{\rm int}$ also tend to be higher with the 75 source sample. A paired sample \textit{t}-test shows that the variation of $\log_{10}T_{\rm int}$ with different constraints on $t_{i+1}-t_{i}$ is significant ($\textrm{\textit{p}-value}<0.01$) at both frequencies. This suggests that the estimates of $\log_{10}T_{\rm int}$ may be affected by (changes in) the source sample as well as observation cadence. Therefore, we refer to the values obtained with the constraint $t_{i+1}-t_{i}<30$~days as upper limits on both the 15~GHz data and the 43~GHz data.\\
\indent{}Additionally, we conducted a D’Agostino and Pearson’s test (DP test) and Shapiro-Wilk test (SW test) on the 15 and 43~GHz $\log_{10}T_{\rm int}$ samples to quantify deviations from a normal distribution. At 43~GHz, both DP and SW tests resulted in \textit{p}-values $>0.1$ for most $t_{i+1}-t_{i}$ constraints. The only exception was the SW test combined with the 30-day constraint, resulting in a \textit{p}-value of $0.0098$. At 15~GHz, only the DP test with a 500-day constraint results in a \textit{p}-value of $>0.05$. All other cases result in \textit{p}-values $<0.0129$, with most less than 0.001. This suggests that while the distribution of $\log_{10}T_{\rm int}$ approximately follows a normal distribution at
43 GHz, there is a significant deviation from normality at 15 GHz.\\
\begin{table*}[t!]
\caption{\label{tab:tint_var_bias}Dependence of $T_{\rm int}$ on data constraints used.}
\centering
\begin{tabular}{cccccc}
\hline\hline
\multirow{2}{*}{$\nu^{\rm o}$}& \multicolumn{5}{c}{$\log_{10}T_{\rm int}\textrm{~[K]}$, $\left<\log_{10}T_{\rm int}\textrm{~[K]}\right>$} \\
\cline{2-6}
 & \multicolumn{5}{c}{$t_{i+1}-t_{i}$}\\
\textrm{[GHz]} & $<30$~days & $<50$~days & $<100$~days & $<500$~days & $<\infty$~days \\
\hline
\multirow{2}{*}{$15$} & $10.73^{+0.83}_{-0.99}$, $10.63^{+0.07}_{-0.07}$ & $10.73^{+0.86}_{-1.01}$, $10.66^{+0.05}_{-0.05}$ & $10.88^{+0.78}_{-0.87}$, $10.84^{+0.04}_{-0.04}$ & $11.13^{+0.73}_{-0.81}$, $11.09^{+0.03}_{-0.03}$ & $11.19^{+0.75}_{-0.80}$, $11.15^{+0.03}_{-0.03}$ \\
 & (75) & (130) & (234) & (350) & (356)\\
\multirow{2}{*}{$43$} & $10.74^{+0.91}_{-1.12}$, $10.68^{+0.13}_{-0.13}$ & $10.85^{+0.94}_{-1.00}$, $10.84^{+0.14}_{-0.13}$ & $10.91^{+0.95}_{-0.96}$, $10.91^{+0.13}_{-0.13}$ & $10.92^{+0.94}_{-0.96}$, $10.91^{+0.13}_{-0.13}$ & $10.92^{+0.94}_{-0.96}$, $10.91^{+0.13}_{-0.13}$ \\
 & (37) & (37) & (38) & (38) & (38)\\
\hline
\end{tabular}
\tablefoot{Values of $\log_{10}T_{\rm int}\textrm{~[K]}$ and $\left<\log_{10}T_{\rm int}\textrm{~[K]}\right>$ for a uniform disk geometry obtained for each of the different constraints on $t_{i+1}-t_{i}$. The first column corresponds observation frequency in units of GHz. Columns 2 to 6 correspond to the constraint on the $t_{i+1}-t_{i}$. The number of sources used in each constraint is given in parentheses under each value.}
\end{table*}
\begin{table*}[t!]
\caption{\label{tab:tint_var_bias_7src}Dependence of $T_{\rm int}$ on data constraints used (fixed sources).}
\centering
\begin{tabular}{c|ccccc}
\hline\hline
\multirow{2}{*}{$\nu^{\rm o}$} & \multicolumn{5}{c}{$\log_{10}T_{\rm int}\textrm{~[K]}$, $\left<\log_{10}T_{\rm int}\textrm{~[K]}\right>$} \\
\cline{2-6}
 & \multicolumn{5}{c}{$t_{i+1}-t_{i}$}\\
\textrm{[GHz]} & $<30$~days & $<50$~days & $<100$~days & $<500$~days & $<\infty$~days \\
\hline
$15$ & $10.73^{+0.83}_{-1.00}$, $10.63^{+0.07}_{-0.07}$ & $10.84^{+0.85}_{-0.97}$, $10.77^{+0.07}_{-0.07}$ & $11.05^{+0.85}_{-0.95}$, $11.00^{+0.08}_{-0.08}$ & $11.32^{+0.77}_{-0.88}$, $11.27^{+0.08}_{-0.08}$ & $11.37^{+0.78}_{-0.86}$, $11.33^{+0.08}_{-0.08}$ \\
$43$ & $10.74^{+0.91}_{-1.13}$, $10.68^{+0.14}_{-0.14}$ & $10.86^{+0.93}_{-1.01}$, $10.84^{+0.13}_{-0.12}$ & $10.93^{+0.95}_{-1.00}$, $10.91^{+0.13}_{-0.13}$ & $10.93^{+0.94}_{-1.00}$, $10.91^{+0.14}_{-0.13}$ & $10.94^{+0.93}_{-0.99}$, $10.92^{+0.13}_{-0.13}$ \\
\hline
\end{tabular}
\tablefoot{The same as Table~\ref{tab:tint_var_bias}, but with the source sample fixed to the 75 sources at 15~GHz and 37 sources at 43~GHz that are common in all constraint sets. The number of sources used in each constraint has been omitted as it is fixed throughout for each frequency.}
\end{table*}
While the data suggests that we are currently cadence-limited, we briefly consider methods that may reduce the uncertainties associated with the variability analysis. The width of the distribution of $T_{\rm int}$ derived from variability is quite large, spanning a few orders of magnitude. A significant contributor to the large scatter is the large uncertainties associated with determining $\tau_{i+1,i}$. Such uncertainties may be reduced if we attempt to fit an e-folding timescale to multiple successive points \citep[as in e.g.,][]{2020MNRAS.495L..27H} or through light curve decomposition, should the observation cadence and number of data points be sufficient. As a simple test, we repeat the analysis by calculating the e-folding timescale from the first and last measurement of segments with a continuous increase in flux density. We find that there are changes of $\Delta\log_{\rm 10}T_{\rm int}\textrm{[K]}=-0.06$ to $+0.01$ for the 15~GHz data and $\Delta\log_{\rm 10}T_{\rm int}\textrm{[K]}=-0.08$ to $+0.02$ for the 43~GHz data. Such variations are approximately an order of magnitude smaller than the uncertainty in $\log_{\rm 10}T_{\rm int}$ of $\sigma_{\log_{\rm 10}T_{\rm int}\textrm{[K]}}\approx1$. It could be that there are multiple, overlapping flares in the light curves of VLBI-resolved cores as well, the superpositon of which are affecting our e-folding timescale measurements. The current data does not have sufficient observation cadence for a robust light curve decomposition. Constraining a representative characteristic timescale per source (instead of using the values from each valid pair) may also help reduce the uncertainties.\\
\indent{}Additional broadening of the distribution of $T_{\rm int}$ may come from systematics in our analysis. Of particular interest are potential source-dependent variations of $T_{\rm int}$. The sources used in our analysis come from a wide range of redshifts, and therefore a wide range of $\nu^{\rm a}$. If there is indeed a frequency dependence of $T_{\rm int}$, we would naturally expect the observed value of $T_{\rm int}$ to vary between sources. As $T_{\rm int}$ depends on the jet parameters of each source, there may be a non-negligible source-per-source variation as well. Given the current limitations, such variations are not apparent. It is possible that larger flares with longer variability timescales are in fact, sufficiently time-resolved, and may point toward two (or more) $T_{\rm int}$ populations. We will investigate the difference of $T_{\rm int}$ between the quiescent state and the larger flares in a separate, follow-up paper.\\
\indent{}The distribution of $T_{\rm vlb}$ and $\beta_{\rm app}$ used in the population analysis of Section~\ref{sec:Tint_pop} is shown in Figure~\ref{fig:Tint_pop_per_source}. Blue circles represent constrained measurements of $T_{\rm vlb}$. The dashed line represents the maximum $\beta_{\rm app}$ for a given bulk Lorentz factor (i.e., Equation~(\ref{eq:max_bapp_Tvlb})). Sources that fall under this line have a viewing angle smaller than the critical angle. The solid envelope corresponds to the expected value of $\beta_{\rm app}$ as a function of $T_{\rm vlb}$ for a bulk Lorentz factor of $\gamma_{\rm j}=40$ at 24 and 43~GHz, while $\gamma_{\rm j}=35$ at 86~GHz, where the value of $\gamma_{\rm j}$ at each frequency was determined based on the maximum $\beta_{\rm app}$ of the sources in the sample. The values of $T_{\rm int}$ found for a Gaussian source are given in Table~\ref{tab:KQW_pop_Tbint}.\\
\begin{table}[t!]
\caption{\label{tab:KQW_pop_Tbint}Population study estimates of $T_{\rm int}$.}
\centering
\begin{tabular}{ccc}
\hline\hline
$\nu^{\rm o}$ & Model & $T_{\rm int}$ \\
$(\textrm{GHz})$ &  & $(\textrm{K})$ \\
\hline
24 & gauss & $1.25^{+1.39}_{-0.30}\times10^{10}$\\
43 & gauss & $3.06^{+3.83}_{-0.73}\times10^{9}$\\
86 & gauss & $4.07^{+3.05}_{-0.55}\times10^{9}$\\
\hline
\end{tabular}
\tablefoot{The $T_{\rm int}$ as derived from a population study of the observed $T_{\rm vlb}$ for a Gaussian source model. Each column corresponds to $\nu^{\rm o}$: the observing frequency in GHz, Model: the source geometry model, and $T_{\rm int}$: the nominal value and limits on $T_{\rm int}$. $T_{\rm int}$ for a disk or sphere may be found by multiplying a factor of 0.56 or 0.67, respectively}.
\end{table}
\indent{}The values obtained from the population analysis in this paper are lower by approximately two orders of magnitude compared to \citet{2019A&A...622A..92N}. We shortly discuss a couple of additional factors which may be biasing our results. $\beta_{\rm app}$ at all frequencies were determined using the maximum apparent speeds of jet components at 15~GHz. \citet{2021ApJ...923...67H} find that the maximum $\beta_{\rm app}$ does have the strongest correlation with the median $T_{\rm vlb}$ of the core (compared to the median $\beta_{\rm app}$ and $\beta_{\rm app}$ of the jet component closest to the core). However, multiple studies suggest acceleration of the jet plasma in the millimeter core region \citep[e.g.,][]{2016ApJ...826..135L,2025A&A...695A.233R}. In this case, $\beta_{\rm app}$ at the core should be lower than that implied by the jet components. To test the affect of this, we multiplied the observed $\beta_{\rm app}$ of each source by a random factor drawn from a uniform distribution between 0 to 1. For a Gaussian source model, we find that the nominal value of $T_{\rm int}$ is now $3.05^{+0.31}_{-0.31}\times10^{10}$~K, $7.22^{+1.24}_{-1.24}\times10^{9}$~K, and $9.19^{+1.22}_{-1.22}\times10^{9}$~K at 24, 43, and 86~GHz, respectively. This is a factor of $\sim2.4$ increase compared to when we have used the maximum $\beta_{\rm app}$ directly (i.e, the values in Table~\ref{tab:KQW_pop_Tbint}). A more realistic analysis may be done by modeling the jet acceleration as a function of distance from the jet base \citep[as in, e.g.,][]{2025A&A...695A.233R}. However, we find it unlikely that offsets of $\beta_{\rm app}$ would fully account for the two orders of magnitude difference.\\
\indent{}Source variability may also affect the value of $T_{\rm int}$. \citet{2006ApJ...642L.115H} found that while $T_{\rm int}$ estimated for a "median-low brightness temperature state" is (for a Gaussian source model) $T_{\rm int}\approx3\times10^{10}$~K, it could increase to $T_{\rm int}>2\times10^{11}$~K when using the maximum observed brightness temperature, with the lower limit coming from unresolved cores. This suggests that there may be two populations of $T_{\rm int}$, with the average/low state close to the equipartition brightness temperature \citep{1994ApJ...426...51R}, and the active/high state near the inverse Compton limit \citep{1969ApJ...155L..71K}.
\citet{2019A&A...622A..92N} assumes a common $\gamma_{\rm j}$ over all sources instead of utilizing per-source $\beta_{\rm app}$ values. Further evaluation of the affect of a distribution of $\gamma_{\rm j}$ \citep[as found in, e.g.,][]{2021ApJ...923...67H} on the analysis of \citet{2019A&A...622A..92N}, as opposed to the use of a common value of $\gamma_{
\rm j}$ over all sources, may also further reduce the offsets between the two methods.
\section{$T_{\rm int}$ of the flaring component in 3C~84}\label{sec:app_3c84_c3}
\citet{2023MNRAS.521L..44H} estimate $T_{\rm int}$ for a flaring component as $T_{\rm int}\sim10^{11}$~K at the peak of the flare. We recalculate this using the updated equations for $T_{\rm int}$ given in this paper (i.e., Equation~(\ref{eq:wyc_tbint_re_0})). We adopt $S^{\rm o}_{\nu^{\rm o}}=15.68\pm1.53$~Jy, $0.5\theta_{\rm FWHM}=0.20\pm0.02$~mas, $\Delta t^{\rm o}_{\nu^{\rm o}}=146\pm5$~days from \citet{2020MNRAS.495L..27H} and $z=0.0176$ from \citet{1992ApJS...83...29S}. Taking $\Omega_{\rm m}=0.315$, $\Omega_{\rm \Lambda}=0.685$, $H_{0}=67.4~\textrm{km/s/Mpc}$ \citep{2020A&A...641A...6P}, we find $T_{\rm int}\approx3.74^{+1.44}_{-0.97}\times10^{10}$~K when assuming a disk geometry and $T_{\rm int}\approx3.94^{+1.51}_{-1.02}\times10^{10}$~K when assuming a spherical geometry. We see that once appropriately accounting for source geometry, the intrinsic brightness temperature of the component is consistent with the equipartition brightness temperature limit of $T_{\rm eq}\approx5\times10^{10}$~K \citep{1994ApJ...426...51R}. It can be inferred that at the peak of the radio flare, the jet component may have been in equipartition between the magnetic and kinetic energy densities. 
The Doppler factor is found to be $\delta\approx1.12\pm0.12$ for a sphere and $\delta\approx0.99^{+0.11}_{-0.10}$ for a disk. As with \citet{2023MNRAS.521L..44H}, we find that the values of $\delta$ are consistent with no Doppler boosting (i.e., $\delta=1$). The values of $T_{\rm int}$ and $\delta$ derived for different geometrical models, as well as for $H_{0}=73.04~\textrm{km/s/Mpc}$ \citep{2022ApJ...934L...7R} are summarized in Table~\ref{tab:3C84_C3_Tbint_delta}.
\begin{table}[t!]
\caption{\label{tab:3C84_C3_Tbint_delta}The $T_{\rm int}$ and $\delta$ of the jet component of 3C~84.}
\centering
\begin{tabular}{cccc}
\hline\hline
$H_{0}$ & Model & $T_{\rm int}$ & $\delta$ \\
$(\textrm{km/s/Mpc})$ &  & $(10^{10}\textrm{K})$ &  \\
\hline
& H2023 & $8.64^{+3.32}_{-2.24}$ & $1.06\pm0.11$ \\
67.4 & sphere & $3.94^{+1.51}_{-1.02}$ & $1.12\pm0.12$ \\
& disk & $3.74^{+1.44}_{-0.97}$ & $0.99^{+0.11}_{-0.10}$ \\
         \hline
& H2023 & $9.36^{+3.66}_{-2.50}$ & $0.97^{+0.11}_{-0.10}$ \\
73.04 & sphere & $4.27^{+1.67}_{-1.14}$ & $1.03\pm0.11$ \\
& disk & $4.05^{+1.58}_{-1.08}$ & $0.92\pm0.10$ \\
\hline
\end{tabular}
\tablefoot{The intrinsic brightness temperature and Doppler factor of the jet component of 3C~84 used for distance estimates. Each column corresponds to $H_{0}$: the Hubble constant used, Model: the source geometry model, $T_{\rm int}$: the value and 1 sigma confidence intervals of $T_{\rm int}$, and $\delta$: the value and 1 sigma confidence intervals of the Doppler factor. The H2023 model corresponds to the results when using the method outlined in \citet{2023MNRAS.521L..44H}. Models sphere and disk correspond to the results obtained when using the revised equations in this paper.}
\end{table}
\end{appendix}
\end{document}